\documentclass[conference]{IEEEtran}

\usepackage[utf8]{inputenc}
\usepackage{amsmath}
\usepackage{amssymb}
\usepackage[usenames,dvipsnames]{xcolor}
\usepackage{tikz}  
\usepackage{graphicx}
\usepackage{fancyhdr}
\usepackage{caption}
\usepackage[hidelinks]{hyperref}
\usepackage{subcaption}
\usepackage{tabularx}
\usepackage[noabbrev]{cleveref}
\usepackage{rotating}
\usepackage{multicol}
\usepackage{multirow}
\usepackage{booktabs}
\usepackage{xcolor}
\newcolumntype{L}[1]{>{\raggedright\arraybackslash}p{#1}} % linksbündig mit Breitenangabe
\newcolumntype{C}[1]{>{\centering\arraybackslash}p{#1}} % zentriert mit Breitenangabe
\newcolumntype{R}[1]{>{\raggedleft\arraybackslash}p{#1}} % rechtsbündig mit Breitenangabe
\usepackage{textcomp}
\usepackage{eurosym}
\usepackage{footnote}
\usepackage{pbox}
\usepackage[flushleft]{threeparttable}
\PassOptionsToPackage{hyphens}{url}\usepackage{url}
\usepackage[deletedmarkup=<deletedmarkup>]{changes}

\makesavenoteenv{tabular}
\makesavenoteenv{table}

\definecolor{darkspringgreen}{rgb}{0.09, 0.45, 0.27}

\if01

	\newcommand{\crev}[1]{{\color{darkgray} [\underline{Comment for Reviewers:} #1]}}
\else

	\newcommand{\crev}[1]{}
\fi

\newcommand\tabrotate[1]{\begin{turn}{90}\rlap{#1}\end{turn}}

\ifCLASSOPTIONcompsoc
  \usepackage[nocompress]{cite}
\else
  \usepackage{cite}
\fi

\usepackage{enumitem}

\newenvironment{packed_item}{
\begin{itemize}[leftmargin=0.4cm]
  \setlength{\itemsep}{1pt}
  \setlength{\parskip}{0pt}
  \setlength{\parsep}{0pt}
}{\end{itemize}}

\newenvironment{packed_desc}{
\begin{description}[leftmargin=0.4cm]
  \setlength{\itemsep}{1pt}
  \setlength{\parskip}{0pt}
  \setlength{\parsep}{1pt}
}{\end{description}}

\newcommand{\seclabel}{security update label}
\newcommand{\seclabelb}{Security update label}
\newcommand{\seclabelu}{Security Update Label}

\begin{document}

%%%%%%%%%%%%%%%%%%%%%%
% TITLE AND AUTHORS
%%%%%%%%%%%%%%%%%%%%%%

\title{\Huge Security Update Labels: \\ Establishing Economic Incentives for  \\ Security Patching of IoT Consumer Products}

\author{\IEEEauthorblockN{Philipp Morgner$^1$, Christoph Mai$^2$, Nicole Koschate-Fischer$^2$, Felix Freiling$^1$, Zinaida Benenson$^1$}
\IEEEauthorblockA{$^1$Department of Computer Science ~ $^2$School of Business, Economics and Society \\ Friedrich-Alexander-Universität Erlangen-Nürnberg (FAU), Germany}
\IEEEauthorblockA{\{philipp.morgner, christoph.mai, nicole.koschate-fischer, felix.freiling, zinaida.benenson\}@fau.de}
}

\maketitle

% Enable page numbers
\thispagestyle{plain}
\pagestyle{plain}

%%%%%%%%%%%%%%%%%%%%%%
% ABSTRACT
%%%%%%%%%%%%%%%%%%%%%%

\begin{abstract}
With the expansion of the Internet of Things (IoT), the number of security incidents due to insecure and misconfigured IoT devices is increasing. 
Especially on the consumer market, manufacturers focus on new features and early releases at the expense of a comprehensive security strategy. 
Hence, experts have started calling for regulation of the IoT consumer market, while policymakers are seeking for suitable regulatory approaches. 
We investigate how  manufacturers can be incentivized to increase sustainable security efforts for IoT products. 
We propose mandatory \emph{security update labels} that inform consumers during buying decisions about the willingness of the manufacturer to provide security updates in the future. 
Mandatory means that the labels explicitly state when security updates are not guaranteed.
We conducted a user study with more than 1,400 participants to assess the importance of security update labels for the consumer choice by means of a conjoint analysis. 
The results show that the availability of security updates (until which date the updates are guaranteed) accounts for 8\% to 35\% impact on overall consumers' choice, depending on the perceived security risk of the product category.
For products with a high perceived security risk, this availability is twice as important as other high-ranked product attributes. 
Moreover, provisioning time for security updates (how quickly the product will be patched after a vulnerability is discovered) additionally accounts for 7\% to 25\% impact on consumers' choices. 
The proposed labels are intuitively understood by consumers, do not require product assessments by third parties before release, and have a potential to incentivize manufacturers to provide sustainable security support. 
\end{abstract}

\IEEEpeerreviewmaketitle

%%%%%%%%%%%%%%%%%%%%%%%%%%%%
% INTRODUCTION
%%%%%%%%%%%%%%%%%%%%%%%%%%%%

\section{Introduction}
\label{sec:introduction}

In 1999, Kevin Ashton coined the term `Internet of Things'~\cite{ashton2009internet} as the headline of a marketing presentation that promoted an idea of utilizing radio-frequency identification (RFID) in supply chains.
By now, Internet of Things evolved into a major technological paradigm: everyday items, household appliances, and mobile devices are interconnected via wireless networks and the Internet.
Gartner \cite{gartner2017} predicts that the majority of IoT devices, 12.9 billion units (63\%), will be installed in the consumer sector by 2020, and thus, consumer products play a prominent role in the expansion of the IoT.

Recent academic and industrial user studies~\cite{DBLP:conf/soups/ZengMR17,mcafee2018,he2018rethinking,DBLP:conf/chi/NaeiniDAC19} document various security concerns regarding the usage of IoT products. 
At least since the denial-of-service attacks against Internet infrastructure by the Mirai botnet \cite{DBLP:conf/uss/AntonakakisABBB17} in 2016, security experts have started to demand regulatory interventions.
``Our choice isn’t between government involvement and no government involvement'', says Bruce Schneier in his testimony before a committee of the U.S.\ House of Representatives~\cite{schneier2016committee}, ``Our choice is between smarter government involvement and stupider government involvement''.
Current policy approaches in the U.S.\ include a  bill  for establishing guidelines for the acquisition of secure IoT products by governmental agencies~\cite{usiotsecact2017}
%https://nakedsecurity.sophos.com/2018/09/13/california-bill-regulates-iot-for-first-time-in-us/
as well as a Californian bill \cite{california-iot-bill} obligating manufacturers to equip IoT devices with reasonable security features.
In the EU, baseline security recommendations for IoT were published by the European Union Agency for Network and Information Security (ENISA)~\cite{enisa2017}.
%  https://www.enisa.europa.eu/news/enisa-news/eu-leaders-agree-on-ground-breaking-regulation-for-cybersecurity-agency-enisa
%In addition, the EU commissioned ENISA with overseeing IoT security policy work and coordinate responses on cyber-incidents  \cite{enisa-agency}.
A task force from academia, industry, and societal organizations proposed a policy for vulnerability disclosure in the EU that also concerns IoT products~\cite{ceps2018}.

The deficient IoT security can be at least partly attributed to missing economic incentives for manufacturers. 
To be successful on the market, manufacturers have to attract consumers and complementers~\cite{anderson2001, anderson2006}.
Consumers reward an early market entry and new functional features, while complementers favor systems that allow easy compatibility with their products.
These demands contradict the security design that usually adds complexity to systems.  
In addition, releasing an innovative product to the market requires many resources, and since resources are finite, they are withdrawn from non-functional features, such as comprehensive security mechanisms \cite{morgner-ndss-diss}.

The missing incentives for securing IoT consumer products originate from the consumers' inability  to compare security properties of different products.
% is a major problem that 
 The concept of an asymmetric information barrier between buyers and sellers, which also affects other properties, such as energy consumption and product quality, is known in the economic theory as `the market for lemons' \cite{akerlof}.
This theory states that consumers are not willing to pay a price premium for something they cannot measure.
%\revision{This applies to security in IoT consumer products: 
%In contrast to the so-called \emph{search} features that can be evaluated by consumers during the buying decision, security is an \emph{experience} feature that can only be discovered during the usage of a product~\cite{doi:10.1086/259630}.}
%consumers are left to trust the claims of manufacturers. Quite often, these claims are rather advertising promises than actual security guarantees. % security vulnerabilities have been disclosed in `secure' products.
% How can a consumer determine the quality of security measures implemented in a product?
 In fact, even manufacturers might not have the complete knowledge about the strength of their products' security \cite{anderson2006}. % as this is not part of their immediate business goals.
Reasons might be a lack of experience in designing Internet-connected technologies or the outsourcing of a product's  security development to original equipment manufacturers (OEMs).

\subsubsection*{Contributions}
Firstly, we propose and examine mandatory \emph{\seclabel{}s}, a novel idea for a regulatory framework that complements ongoing regulation efforts. We do not call for security testing and certification to keep ``insecure'' products off the market. Instead, we explore to which extent market forces can be utilized to elicit manufacturers to  sustainably support their products' software with security updates. \seclabelb{}s enable an  informed choice regarding security properties of IoT consumer products. 
They transform the asymmetric information about the manufacturer's willingness to provide security updates into two intuitively assessable and comparable product attributes: \emph{availability period}, i.e., for how long the manufacturer guarantees to provide security updates (e.g., `until 12/2016'), as well as \emph{provisioning time}, i.e., within which timeframe after a vulnerability notification a security patch is provided (e.g., `within 30 days'). These labels are inspired by established regulations, such as energy labels.

Secondly, we empirically examine the impact of \seclabel{}s on the consumers' choice. 
Although security patching is discussed by experts as one of the most effective countermeasure against insecure IoT devices, the impact of guaranteeing security updates on the consumers' decisions has not been empirically assessed so far.
We conducted a user study with more than 1,400 participants that measured the relative importance of the availability period and provisioning time of security updates for buying decisions. To this end, we used conjoint analysis, a well-established method in marketing research \cite{green2001,wittink1994}, which has also been used in courts to calculate damages of patent and copyright infringements \cite{goodcounsel-conjoint}.
In a nutshell, a number of fictitious product profiles, each described by a set of attributes, is shown to respondents in multiple iterations. 
They are asked which of the presented products they would prefer to buy (with the option to refuse buying any of the products).
Based on these choice results, conjoint analysis determines a preference model that measures the relative importance and utility of each attribute.

% to the best of our knowledge

%With an interdisciplinary team consisting of researchers from the domains of IoT security, human factors in security, marketing research, and psychology, we conducted a user study with more than 1,400 participants that measured the relative importance of the availability period and provisioning time for security updates for buying decisions.

\subsubsection*{Study Results}
We found that the guarantee of providing security updates has a high impact on buying decisions.
We examined two product categories, one with a high and one with a low perceived security risk.
%We discovered differences between both categories in the importance of the availability period for security updates.
Among all assessed product attributes, the availability period of security updates was the most important one:
For the product with the high perceived security risk, its relative importance on the overall consumers' choice of 31\% is at least twice as high as the importance of  other attributes.
For the product with the low perceived security risk, availability had a lower relative importance of 20\% for the consumers' choice. %, which was slightly higher than for other high-ranked attributes.
%However, both relative importance are the highest among the importance of all assessed product attributes.
Additionally, consumers prefer a shorter provisioning time (10 days) over a longer provisioning time (30 days), and dislike longer provisioning times for products with a high perceived security risk.
Demographic characteristics play a minor role, while the sensitivity for security risks has an impact on the consumers' choice. % in terms of the \seclabel{}.

% Implications
%\revision{In summary, we propose and empirically analyze security attributes for IoT consumer products that are recognized by consumers and do not require third-party product testing.}{}
With this work, we address policymakers and security researchers that are seeking for promising directions to foster sustainable security efforts for IoT consumer products. %,

\section{Background and Related Work}
\label{sec:background}

We provide background and related work on product labeling and conjoint analysis in this section.
%\zina{would be good to have a subsection on security and privacy labels, even if it is going to be small --> as  reviewer I would expect this; here also attestation can be mentioned, because it also results in labeling (I think)}

\subsection{Product Labeling}
\label{sec:background:labeling}

Product labeling is used in many countries to inform consumers about intangible features of products and to enable product comparison during buying decisions.
The Federal Trade Commission (FTC) issues product labeling policies in the USA, while each member state of the EU runs its own institution that enforces regulations defined by the EU Commission.
%
%In the past, a number of product labeling policies have been introduced to reduce information asymmetries between manufacturers and consumers.
Prominent examples are energy labels. % that inform about the energy consumption of particular products.
%\revision{In the USA, designated electronic products, e.g., light bulbs, televisions, and household appliances, must be tagged with labels 
%%as exemplarily depicted in \Cref{fig:label:ftc-energy}, 
%that show the energy consumption and the estimated annual operating costs. 
%These labels have been introduced by the FTC with the Energy Labeling Rule \cite{ftc-energy}
%%, a part of the Energy Independence and Security Act of 
%in 2007 \cite{eisa}.}{
In the USA, energy labels were introduced in 2007 \cite{eisa} by the FTC and show the energy consumption and the estimated annual operating costs.
%  with the Energy Labeling Rule \cite{ftc-energy}
In 2010, The EU followed with a similar approach by introducing the Energy Efficiency Directive \cite{eu-lex-energy}. % that demands energy labels
%, as exemplarily shown in \Cref{fig:label:eu-energy}, 
%for certain product categories.
%The EU energy labels are part of the efforts to foster energy-efficient products with the objective to reduce the overall energy consumption of the EU by 20\% until 2020~\cite{eu-energy-dir}.
%
Prior research on the effectiveness of energy labeling \cite{winward1998cool,waide2001monitoring,BSE:BSE522} concluded that consumers are aware of these labels, understand them, and that energy labels influence consumers' buying decisions.

% LIFETIME LABELS
In 2017, the German government~\cite{german-lifetime-label2017} evaluated an idea of lifetime labels on electronic products.
%The idea behind their label was to evaluate a way to endorse transparency and sustainability in buying decisions.
Their label design showed a color-gradient lifespan between 0 (red) and 20 (green) years.
In a user study with a representative sample, discrete-choice experiments (but not conjoint analysis) simulated online shopping scenarios. %purchasing situations in
The results showed that while the lifespan attribute was recognized by consumers, its impact on  buying decisions was less than the impact of other product attributes, e.g., price and brand.
Their label did not concern security features, but the functional lifespan of a product. %  for which the product is intended to work properly. solely lifetime 

%have been facilitated that 

% PRIVACY LABELS
\subsection{Security \& Privacy Labels and Regulatory Approaches}
\label{sec:background:privacy-labels}

In the academic research, the adaption of product labels for privacy information was examined  in user studies. 
%Especially the usefulness of privacy labels has been investigated by means of user studies. %These studies did not involve conjoint analysis. % attracted attention in the past decades. 
Kelley et al.\ \cite{DBLP:conf/soups/KelleyBCR09, DBLP:conf/chi/KelleyCBC10} investigated whether food nutrition labels can be adapted to make privacy policies of websites more understandable.
Tsai et al.\ \cite{DBLP:journals/isr/TsaiECA11} evaluated whether consumers would pay a higher price for a product offered by an online shop with a strict privacy policy as compared to a less privacy-protecting shop. % if it is 
%\zina{what conclusions did they make form they studies? what were the results? are they encouraging for us? can we learn something from them?}
Their results suggest that consumers are willing to pay a price premium for higher privacy  if privacy information is salient and understandable.
%\philipp{Add Emami CHI paper}

Independently and concurrently to our work, Emami-Naeini et al.\ \cite{DBLP:conf/chi/NaeiniDAC19} developed  a security and privacy label for IoT consumer products.
In contrast to our proposal, their label includes ratings that require third-party product testing before release. 
They tested their label in an interview study with 24 users and a survey with 200 respondents. 
Emami-Naeini et al.\ did not conduct a conjoint analysis but directly asked the users to rate the importance of security and privacy on their buying decisions. 
They concluded that importance of security and privacy depends on the product category: whereas they are important when buying a home camera or a smart thermostat, they are not important when buying a smart toothbrush. 
We found a similar effect in our study.
Our and their studies complement and validate each other's results using different methods.

Mandatory security update labels represent a possible approach to regulate the IoT product market with regard to security. 
Chattopadhyay et al.~\cite{chattopadhyay-2019-single-item} consider this economic problem in more depth and analyze the impact of various regulation strategies on consumers' behavior.

\subsection{Conjoint Analysis}
\label{sec:background:conjoint}

Conjoint analysis is one of the major methods to measure the impact of product attributes on the consumers' buying decisions~\cite{gilbride2008}. 
%Marketing research provides a number of methods to measure the impact of product attributes on the consumers' buying decisions.
%One of the major data analysis methods is conjoint analysis \cite{gilbride2008}. %, proposed in the 1970s by Paul Green and colleagues \cite{green1978, green1990}.
%They based their methodology on conjoint measurements to decompose consumers' overall judgments into separate and comparable utility scales.
%This technique was originally developed by Luce and Tukey \cite{luce1964} to provide measurement scales between dependent and independent variables.
 The basic idea of conjoint analysis is that respondents are asked to state their preference for buying fictitious products.
%These fictitious products are also denoted as profiles. 
The product profiles are described by a limited set of attributes, e.g., size, color, and price. 
All further attributes of the product are assumed to be constant.

There are different types of conjoint analysis. 
 Among them, choice-based conjoint (CBC) is used in 79\% of the conjoint surveys \cite{orme2014}. %in conjoint data collection that %is considered the de-facto standard and 
In CBC, which we use in this work as well, the respondents receive multiple (usually randomly generated) subsets of 3 to 5 product profiles (so-called choice sets), of which they select the most desirable product.

%In each choice set, there exists also a `no-choice' alternative.
%The responses are collected and analyzed with conjoint analysis software.

% Event. später: ,carroll1969,kruskal1965,young1969
Considering the overall preference (i.e., combination of all buying decisions) as dependent variable and the attributes of the product as independent variables, a conjoint analysis assesses the relative importance of product's attributes. % derived from the overall preference score of a product profile consisting of two or more attributes.
For example, relative importance of the attribute `color' for buying decisions can be assessed.
Conjoint analysis also evaluates the importance of the different characteristics of a single attribute, e.g., whether the change of a product's color would have positive or negative effects on the consumers' choice.

In the past decades, conjoint analysis has been applied to numerous commercial projects \cite{voleti2017} and is by far the most widely-used methodology in marketing research to analyze consumer trade-offs in buying decisions \cite{green2001}.
Conjoint analysis is also used in other areas, e.g., to assess the patients' preferences in the healthcare sector \cite{ryan1997}. 
%
% https://blog.ams-litigation.com/court-upholds-use-conjoint-analysis-classwide-damages
Furthermore, it is a recognized methodology to calculate damages of patent and copyright infringement  in court cases \cite{goodcounsel-conjoint}.
% Apple vs Samsung: 
% http://www.sawtoothsoftware.com/download/apple_v_samsung_conjoint_analysis.pdf
% https://www.theverge.com/2014/4/8/5590278/heres-how-apple-figured-out-the-extra-2-19-billion-it-wants-from
A famous example was Apple's \$2.5 billion law suit against Samsung, in which Apple estimated the financial damages of the alleged patent infringement based on conjoint analysis~\cite{sawtooth-apple}.

% LABELS AND CONJOINT
Conjoint analysis has also previously been used to investigate the effects of product labels \cite{pelsmacker2005,drewnowski2010,hieke2012} on the consumers' choice.
Sammer and Wüstenhagen \cite{sammer2006} analyzed the impact of energy labels on the buying decisions concerning light bulbs and washing machines of Swiss consumers.
%Pelsmacker et al.\  \cite{pelsmacker2005} investigated the influence of fair-trade labels on coffee buying decisions of Belgian consumers. 
%Further work \cite{drewnowski2010,hieke2012} used conjoint analysis to evaluate the influence of food nutrition labels on the consumer choice. 
%
However, we are the first to use conjoint analysis to assess the importance of a security-related label.

\section{Security Labels for Consumers}
\label{sec:label}

Inspired by the success of existing product labels, we propose a label that enables users to compare security properties during buying decisions. 
We present the idea of a regulatory framework that accompanies the label, and  discuss concerns that finally motivate the user study.

\subsection{Security Scales for Labeling}
%label{sec:label:metrics}

Learning from the success of the energy labeling initiatives, we asked  how we can use a similar approach for security.
First of all, an appropriate scale to measure security properties is required.
We need a security scale that 
\begin{enumerate}
\item can be intuitively understood by consumers, even if they have no security expertise;
\item enables them to easily compare products, as comparison lays the foundation for the  choice between products;
\item and finally, does not require third-party product testing for market release.
\end{enumerate}

The last requirement is based on the following considerations: 
Third-party testing is a long and costly procedure that might considerably delay the release of a new product.
This involves the danger that manufacturers would choose testing laboratories that perform a relaxed and fast evaluation \cite{DBLP:journals/ieeesp/MurdochBA12, leverett2017standardisation}, which again could lead to a false sense of security. %Moreover, any changes in the product will requitre re-attestation.

%As the term `security metric' has different meaning depending on the context, we are looking for security metrics that could be applied to measure the level of security in IoT products for consumers.
 %
In prior work, a number of security scales has been proposed that could be applied to IoT products.
Many of them (e.g.,~\cite{boyer2008, DBLP:journals/comcom/LaiH07}) are based on the Common Vulnerability Scoring System (CVSS) \cite{cvss2007} and used  to categorize the seriousness and impact of existing security vulnerabilities.
%by incident response teams, security service providers, and researchers
%%CVSS categorizes the seriousness of known vulnerabilities on system level in three metrics.
%The major scale of CVSS is the so-called base score ranging from 0 (low severity) to 10 (high severity), which is calculated taking the exploitability and impact of a vulnerability into account.
%%
%The set of proposed CVSS-based security metrics \cite{holm2012} includes weakest link metrics (i.e., security is as strong as the vulnerability that requires the least effort to exploit) \cite{wang1997}, vulnerability exposure metrics (i.e., the total number of vulnerability days) \cite{boyer2008}, number of vulnerabilities metrics (i.e., sum of all vulnerabilities), and others \cite{DBLP:journals/comcom/LaiH07,DBLP:conf/IEEEares/TupperZ08,DBLP:conf/dasc/BonillaCBC17}.
%% 
%Although CVSS is used by practitioners in industry to categorize the vulnerability of  systems, a security metric based on the CVSS does not seem to be reasonable for IoT consumer products.
Although CVSS can serve as an indicator of future security properties, it cannot solely measure the current level of product security, as it is based on past vulnerability records. 

The time-to-compromise (TTC) \cite{leversage2008} scale originated from the concept of the working time required to break a physical safe.
In terms of IoT consumer products, this metric could measure the time it takes to break the security mechanisms of a product.
According to our criteria,  TTC is not applicable as it  requires a third party to assess the product's security.

% quality of software and/or hardware architecture. 
%In contrast to drilling a hole in a physical safe, there is no standard procedure on how to break the security measures of an IoT consumer product.

%classification of an IoT consumer product 

A security scale might also show levels of a security certification scheme. % as established in the industrial and military sector.
%While security certification can help to prevent security vulnerabilities in the design and implementation of the products, it is not suitable for our purposes.
However, besides the need for a third party, security certification is not suitable to communicate security levels to consumers, as it might be misleading: 
Whereas consumers may assume that the whole product is certified, in reality only a subset of the components might be certified~\cite{DBLP:journals/ieeesp/MurdochBA12}.

%, which verifies that the product is compliant to a set of security requirements, 

\subsection{\seclabelu{}s}
\label{sec:label:idea}
\label{sec:label:patching}

We conclude that, to the best of our knowledge, there are no suitable approaches to communicate the security level of an IoT product to consumers.
And even if manufacturers would implement comprehensive security measures, security flaws in IoT products cannot be fully prevented.
% Excluded: DBLP:conf/hase/MalaiyaD98
Prior research \cite{DBLP:journals/compsec/AlhazmiMR07,jones2011} concluded that well-engineered code has an average defect rate of around 2 defects per 1,000 lines of code.
%Each larger piece of code is imperfect and contains a number of defects.
%Some researchers even claim, a state of perfect security will never be reached as the system would lose its usefulness \cite[p.61]{DBLP:series/ccn/Kizza17}.
%Hence,  security efforts should be seen as the continuous support to repair such defects as soon as they are disclosed. 
%how can we continuously support the repair of such defects as soon as they are disclosed?
% with best-effort security
If we accept the possibility of security vulnerabilities even in well-designed systems, the best approach would be to continuously support the repair of such defects as soon as they are disclosed.
We propose a regulatory framework that demands brand-giving manufacturers to define an \emph{update policy} for each  IoT consumer product with the following properties: %newly released
%The core properties of this update policy comprise following information:  

%packed_desc

%\philipp{(5) Clarify use of other features}

\begin{description}[leftmargin=0.4cm]
\item[Availability period:] 
The availability of security updates determines the absolute timeframe in which the manufacturer ensures the patching of security vulnerabilities in the product's software.
In other words, it defines until which date (for example: `12/2024') the manufacturer contractually warrants  to provide security updates.

\item[Provisioning time:] 
When a security vulnerability in  the software of an IoT consumer product was reported, the manufacturer has to investigate this issue and patch the software if needed.
The update policy defines the maximum timeframe (for example: `30 days') within which the manufacturer guarantees to provide software security updates.

\end{description}

Both attributes must be printed as a \emph{\seclabel{}} on each adequate product such that consumers can compare this  information when making a buying decision.
The label content does not need to be authorized by a third party before the market release, similarly to the mandatory energy labels.
If a manufacturer refuses to guarantee security updates, the label should explicitly display `no security updates guaranteed' or a similar phrase. %}{state this}.
%\revision{This label transforms the security update attributes from experience attributes (i.e., they can only be assessed after purchase) into search attributes (i.e., they can be evaluated before purchase).}{}

% `availability period' and `provisioning time'

\subsection{An Idea for a Regulatory Framework}
\label{sec:label:legislation}

Following the example of the legislations for energy labeling \cite{eu-lex-energy, ftc-energy}, market surveillance and consumer protection authorities should supervise the implementation of the \seclabel{}s, and conduct promotional and educational information campaigns in the introduction phase.
We propose that each applicable product displays the label on the packaging, such that it can be considered and compared during buying decisions, and on the device itself to inform the consumer about the guaranteed availability of security updates after deployment.
These labels should be mandatory for each consumer product that is able to directly or indirectly (e.g., over Bluetooth) connect to the Internet. 
%A direct connection means that the product supports the Internet protocol (IP). 
%An indirect connection means that the product supports a non-IP protocol (e.g., Bluetooth), while an Internet-connected bridge node translates protocol messages into IP traffic and vice versa.}
The liability should be enforced only between the brand-giving company that is responsible for the definition of the update policy, and the buyer of the product. 
All further interactions between the brand-giving company and OEMs or other involved third parties should be regulated by the market.

The vulnerability disclosure can be implemented in many ways.
An approach might be to set up a public vulnerability reporting platform. 
This platform could ensure the documentation of the reported vulnerabilities and would act as an information channel where the manufacturer announces the current state of the vulnerability handling to the affected consumers and policy-enforcing entities.
The design of such a reporting platform could follow the proposal of the Centre for European Policy Studies \cite[p.56]{ceps2018} and is out of scope of this paper. 
Procedures could be implemented based on the established standards for responsible vulnerability disclosure, e.g., ISO/IEC 29147 \cite{iso29147}, and vulnerability handling, e.g., ISO/IEC 30111 \cite{iso30111}. %, and also follow further best practices on vulnerability disclosure, e.g., by ENISA \cite{enisa-vuln-2015}. 
When a suspected security vulnerability is found, the reporting entity files a vulnerability report via this platform, which in turn informs the affected manufacturers.
After receiving the vulnerability report, the time-to-patch clock starts and the manufacturer investigates whether the vulnerability can be reproduced. %and identified with internal sources
If the manufacturer concludes that the reported vulnerability is an actual security flaw, a security patch shall be developed and provided within the guaranteed provisioning  time.
We propose that consumers have a right to compensation in the following cases:

\begin{packed_item}
\item The manufacturer does not provide a required security patch within the guaranteed provisioning time.
\item The manufacturer provides a security patch, but the patch does not fix the bug, introduces other security problems, or has serious effects on the performance of the product.
\end{packed_item}

For the cases of disputes about the effectiveness of provided updates or whether a bug requires a security patch, policymakers should establish an entity that enforces accountability, judges the claims of the consumers, protects vulnerability reporters, and has the power to sanction manufacturers, similarly to the sanctions imposed by the General Data Protection Regulations (GDPR) in the EU~\cite[Art.58]{eu-gdpr}.

% \zina{any examples or existing ones (in other areas), or a ref to a proposal?}
%These authorities can use processes similar to the General Data Protection Regulations (GDPR) in the EU, which can fine businesses that violate the privacy regulations \cite[Art.58]{eu-gdpr}. 
% https://gdpr-info.eu/art-58-gdpr/

%In case the manufacturer is not able to act according to its update policy, i.e., the vendor cannot fix the security vulnerability within the self-defined period of time, then the consumer should be able to claim compensation.
%If the manufacturer is able to patch the security flaws within the self-defined period of time, then the consumers cannot claim compensations for the security flaws since they made an informed decision when buying these products.

%\philipp{(7) Address problem of vulnerability reporting and accountability}
%\philipp{(8) Discuss that manufacturers are able to define update policies}

\subsection{Concerns towards \seclabelu{}s}
\label{sec:label:concerns}

The proposal of \seclabel{}s might raise the following concerns.
\subsubsection{Ineffectiveness}
Some security vulnerabilities cannot be patched with updates. % because they are fundamental or hardware flaws?
For example, a security flaw in the specification of an interconnected system might demand changes in other components that are not maintained by the manufacturer, or the hardware platform of the affected product cannot support the patched software due to memory or computational power constraints.
In this case, the proposed label strengthens consumer rights as the consumer is entitled to compensation.

\subsubsection{Misuse}
Manufacturers might be motivated to spend even less resources on security of their products before releasing them. 
They might decide that they always can patch the product within a certain timeframe, which means that they simply could outsource the debugging of their products to the consumers.
We believe that such behavior would damage the user acceptance and the brand image.
Furthermore, this practice would  lead to a high pressure on the manufacturers to deliver numerous security patches within limited time.
In another scenario, manufacturers might try to transfer the liability regarding their products to offshore companies.
These scenarios should be considered when defining the legislation.
%From our point of view, manufacturers are more likely to cover potential loss through liability insurances that already cover warranty and malpractice claims. 
%
%Situations of disagreement between consumer and manufacturer can arise about whether a reported problem is a security vulnerability, or whether a released patch is effective.
%In these cases, the effectiveness of the \seclabel{}s depends highly on the entity that has the expertise to judge about the security-criticalness of a bug, or whether a patch addressed the security problem in a sustainable manner.
%There is the possibility to install an entity that judges on conflicts as the jurisdictional systems routinely rule over issues, whose in-depth understanding requires expert knowledge.
%Furthermore, independent experts and security researchers can be hired on a contract basis to assess claimed vulnerabilities and their patches.

\subsubsection{Low User Acceptance}
The \seclabel{}s could fail as they might not have the expected effect on consumers' buying decisions.
Prior user studies \cite{DBLP:conf/chi/VanieaRW14,DBLP:conf/chi/VanieaR16,DBLP:conf/soups/ForgetPTACCEHT16,DBLP:conf/soups/MathurC17} outline that consumers tend to be reluctant towards the installation of updates.
This behavior results from a lack of clarity about the usefulness of updates as well as from negative update experiences in the past, such as unwanted changes in user interfaces or in functionality.
% that occurred from unwanted changes in the user experience, e.g., the remodeling of a graphical user interface.
In consequence, the attitude towards security updates is affected as users typically do not differentiate between different types of updates. 
Therefore, \seclabel{}s could have a low user acceptance.

Potential moral hazard \cite{pauly1968} could also lead to a low user acceptance.
In our context, this means that users might not be willing to pay a price premium to protect against security vulnerabilities that will not affect them.
An illustration are the attacks by the Mirai botnet \cite{DBLP:conf/uss/AntonakakisABBB17}, in which thousands of IoT consumer products deployed in Latin America attacked US-based Internet services.
In this case, why would a Latin-American consumer pay a price premium for a security update guarantee that protects US businesses?

%Nonetheless, consumers are concerned about security, independently from a potential moral hazard. 
%Recent surveys~\cite{DBLP:conf/soups/ZengMR17,mcafee2018} show that security is indeed one of the most important consumer concerns towards IoT products. 

%We consider the concerns of ineffectiveness and potential misuse are out of scope, as they depend on the legislation and  business decisions of particular manufacturers.
The concerns of ineffectiveness and potential misuse depend on the legislation and  business decisions of particular manufacturers. We leave the investigation of these concerns to future work. In the following, we investigate the concern of low user acceptance by means of a user study. 
%Investigating these concerns requires legal and business expertise and is left to future work. 
%Thus, to examine the user acceptance of \seclabel{}s, we conduct a user study that is outlined in the following. 

%%%%%%%%%%%%%%%%%%%%%%%%%%%%
% USER STUDY
%%%%%%%%%%%%%%%%%%%%%%%%%%%%

\section{Concept of User Study}
\label{sec:study}

%\revision{Although the concept of \seclabel{}s seems to be reasonable, there is no evidence of their acceptance by  consumers.}{}
%To investigate whether such labels would indeed impact the buying decisions for IoT consumer products, we conducted a user study. 
%\revision{To investigate this issue, we conducted a user study with the objective to assess whether mandatory \seclabel{}s have the potential to be an important criterion in consumer decision making when buying an IoT product.}{}
If \seclabel{}s turn out to be important for consumers' buying decisions, this would create economic incentives for manufacturers to guarantee the timely patching of security vulnerabilities in their IoT  products.
Therefore, we consider the following research questions:

%{Therefore, we conducted a user study that assessed}

\begin{packed_desc}

\item[RQ1] What is the relative importance of the availability period and provisioning time for security updates for buying decisions compared to other product attributes?

\item[RQ2] Are there differences in the relative importance of the availability period and provisioning time for security updates between products with a high perceived security risk compared to products with a low perceived security risk?

\item[RQ3] Are there differences in the relative importance of the availability and provisioning time for security updates according to demographic characteristics of the consumers?

\item[RQ4] Are there differences in the relative importance of the availability and provisioning time for security updates depending on security behavior intentions, privacy concerns, and security risk perception of the consumers?

\end{packed_desc}

% description
%packed_desc

In the following, we investigate these research questions for German consumers by means of a user study.
%Hereby, we conduct the user study on the German market, which is regulated by national and European policies.
Germany has the largest consumer market within the EU, and the fourth largest consumer market worldwide  after USA, China, and Japan~\cite{worldbank-data}. 
% https://data.worldbank.org/indicator/NE.CON.PRVT.CD?year_high_desc=true

% http://www.euromonitor.com/consumer-electronics-in-germany/report
% https://www.bitkom.org/Presse/Presseinformation/Umsatzanstieg-im-Markt-der-Unterhaltungselektronik.html 
% https://www.statista.com/outlook/251/137/consumer-electronics/germany#

\subsubsection*{Structure of the User Study}
\label{sec:study:structure}

We utilize conjoint analysis, as this method is well suited for our objectives (cf.\ \Cref{sec:background:conjoint}): We aim to determine the influence of the availability period and provisioning time attributes on consumers' choices.
This includes whether these attributes are desired at all (i.e., do consumers care about the availability of security updates?), and 
which attribute levels are more attractive (i.e., do consumers favor short provisioning time or long availability periods?). 

%Our research questions relate to explicit trade-offs between measurable attributes as we 

To answer the research questions, we needed to choose product categories that differ in the perceived security risk.
We decided on two product categories as this number is sufficient to answer the research questions: one with a high perceived security risk as well as one with a low perceived security risk.

\begin{figure}[t]
        \centering
       \includegraphics[trim=0 0 0 0, clip, width=0.48\textwidth]{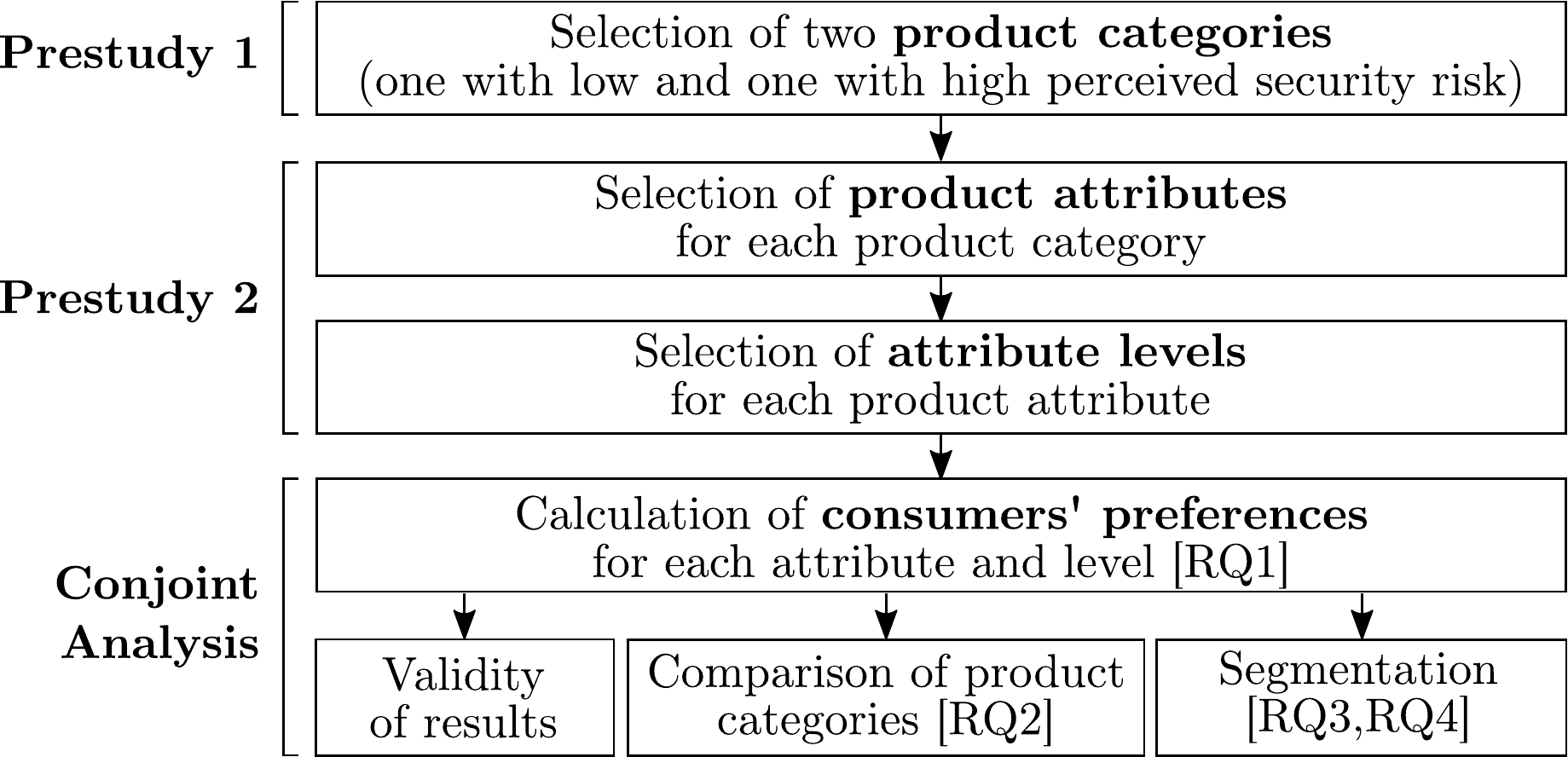}
        \caption{Structure of user study.}
	\label{fig:study-structure}
    \end{figure}

The user study followed a three-stage approach as shown in \Cref{fig:study-structure}:
In the first stage (Prestudy 1), two suitable product categories were selected.
In the second stage (Prestudy 2), we determined the most important product attributes and their levels for each of the two product categories. 
In the third stage (Conjoint Analysis), we assessed the consumers' preferences (RQ1), comparing the attributes of the  \seclabel{} with other important product attributes.
Finally, we validated the preference model, compared the product categories (RQ2), and performed a segmentation analysis (RQ3, RQ4).

% Figure: User Study Structure

\subsubsection*{Ethics and Recruitment}
\label{sec:study:ethics}

The study design was approved by the data protection office of our university. %\footnote{The ethical review board of the authors' institution only considers medical studies.}. 
%\revision{All survey answers were associated with pseudonyms that did not provide any information about the identity of the respondents.}{}
All data was processed in accordance with the German data protection laws and all survey answers were pseudonymized. 
The online surveys were hosted on a web server that is provided by our university, and secured such that only authorized entities have access to the collected data.
The respondents for the online surveys were recruited at an online crowdworker platform, as prior work showed that such samples are appropriate for security research \cite{redmiles-mturk}.
%After examining a number of such platforms regarding functionality, crowd demographics, working conditions, and costs, 
 We used the Clickworker.de platform~\cite{clickworker}, which claims to have the largest crowd of German-speaking workers.
For all online surveys, we selected the respondents with following characteristics: all genders, age between 18 and 65, and Germany as country of residence.
The crowdworkers were paid according to the German minimum wage of \euro{}8.84 per hour. 

%For example, in Prestudy 1, the average time to answer the surveys was estimated through test runsto be 10 to 12 minutes. Thus, we paid each crowdworker a reward of \euro{}1.80 for 12 minutes, adding up to overall cost of \euro{}3.00 per respondent including platform fees and taxes. 

%example, the through test runs, we estimated the average time to answer the surveys to be 10 to 12 minutes.
%Thus, we paid each crowdworker a reward of \euro{}1.80 (reward equivalent for 12 minutes), adding up to overall cost of \euro{}3.00 per respondent including platform fees and taxes. 

\subsubsection*{Translation of Psychometric Scales}
\label{sec:study:translation}

As we run our surveys with German-speaking respondents, we translated all items of utilized English psychometric scales.
These scales measure, e.g., privacy concerns~\cite{dinev-hart-2006} or security behavior intentions \cite{DBLP:conf/chi/SawayaSCKNY17}.
To ensure a reliable translation, we utilized a methodology proposed by Venkatesh et al.\ \cite{viswanath-2012}. %that conveys the meaning and also the statistical properties of the psychometric scales
Three bilingual domain experts translated the English scales into German individually. 
Then, the experts discussed differences in their translations and agreed on a single final version. 
Finally, three other persons (an English native speaker, a professional translator, and a German who lived for several years in the UK) retranslated the German scales back into the original language. 
Through verifying that the original scales matched the retranslated scales semantically, the translation was considered successful. 

%For replication purposes, we provide further details of this study in \Cref{app:replication}.

\subsubsection*{Statistical Data Analysis}
\label{sec:stat}

We denote by $\mu$ the mean value, and by $\sigma$ the standard deviation. 
To assess the practical meaning of the statistical results, we report effect sizes \cite{cohen-1988}: 
For unpaired $t$-tests, the absolute value of $d<0.5$  is considered small, $d$ between 0.5 and 0.8 medium, and $d>0.8$ large effect. 
For paired $t$-tests, effect size $d_z$ is interpreted identically to $d$.
Cramer’s $V$ measures effect sizes for ${\chi}^2$ tests, and $r$ for ANOVA\footnote{$r$ denotes effect size for one-way independent ANOVA according to Field~\cite[p.~472]{field2013-spss} and is calculated as $\sqrt{\eta^2}$.}.
Values around 0.10 indicate a small, 0.30 a medium, and 0.50 a large effect~\cite{field2013-spss}.

%$0.5 \leq d \leq 0.8$
%\cite{cohen-1988,field2013-spss}
%%%%%%%%%%%%%%%%%%%%%%%%%%%%
% SECURITY RISK
%%%%%%%%%%%%%%%%%%%%%%%%%%%%

\section{Preliminary Studies}

\subsection{Prestudy 1: Selection of Product Categories} 
\label{sec:prestudy:categories}

The objective of this prestudy is to identify two categories of IoT products according to the following criteria:
\begin{packed_desc}
\item[C1] 
Both categories should differ significantly in their perceived security risk.

\item[C2]  
Both categories should be similar concerning other central product attitudes and their purchase intentions: 
\emph{Attitude} towards product category (in terms of favor, likability, pleasure) \cite{martin-2005},
\emph{involvement} with product category (in terms of, e.g., fascination, excitement)~\cite{mathwick-2004},
\emph{consumption motive} (hedonistic or utilitarian) \cite{wakefield-2003},
\emph{desirability} to possess products of this product category \cite{lei-2012}, and
\emph{purchase intention} for products of this product category \cite{mackenzie-1986}.
The items of these scales are reported in \Cref{app:prestudy1:scales}.

\end{packed_desc}

%associated with this product category
%enumerate

%As a result, we created a survey that presented eight categories of popular IoT consumer products (

%Both categories should rather not differ concerning central product attitudes and purchase intentions that are often used in marketing research on consumer preferences

\subsubsection*{Perceived Security Risk Scale}
\label{sec:study:prestudy:scale}

To distinguish between product categories with a high and low perceived security risk (criterion C1), we needed a scale that measures the perceived security risk associated with IoT consumer products.
After an extensive literature review, we concluded that there is no scale that sufficiently fits our purpose.
Declined candidates \cite{DBLP:journals/ijmms/FeathermanP03,dinev-hart-2006,luo2010} comprised scales that measure security and privacy risks in e-commerce settings.
However, because IoT products may have adverse effects on the physical world, their security risks are fundamentally different.

Thus, we developed a perceived security risk scale for IoT consumer products using a similar methodology as proposed by Davis \cite{DBLP:journals/misq/Davis89}. 
In the first step, we defined the concept of the perceived security risk in IoT products. 
Perceived risk is defined as the customers' perceptions of uncertainty and unfavorable consequences concerning a product or a service \cite{kushwaha2013}.
In the context of security, uncertainty means the probability of a security incident, while consequences are  the loss caused by such an incident.
We decided to measure only consequences with our scale. We think that it is very difficult for non-experts to determine the probability of a security incident associated with a particular IoT product category, because they would need to assess the quality of the product's security measures as well as the attractiveness of the product  for attackers.
On the other hand, the assessment of consequences of a security incident requires knowledge about the deployment and utilization of the product.
Usage scenarios are known to consumers, and therefore, they can imagine potential consequences.
As a result, we defined that \emph{perceived security risk for an IoT product exists if security vulnerabilities in this product are perceived to lead to negative consequences for the user}.

%Security is the ``absence of unauthorized access to, or handling of, system state'' \cite[p.23]{DBLP:journals/tdsc/AvizienisLRL04}, whereas a system is defined as an ``entity that interacts with other entities, i.e., other systems, including hardware, software, humans, and the physical world with its natural phenomena'' \cite[p.12]{DBLP:journals/tdsc/AvizienisLRL04}.
%Furthermore, security is a combination of the following attributes \cite[p.13]{DBLP:journals/tdsc/AvizienisLRL04}: 
%\begin{itemize}
%\item Confidentiality: absence of unauthorized disclosure of information 
%\item Integrity: absence of improper system alternations
%\item Availability: readiness for correct service
%\end{itemize}
%
%A security incident could lead to the loss of confidentiality, integrity, or availability.
%%We defined perceived risk as a concept consisting of two basic components: uncertainty and consequences \cite{kaplan1974,kushwaha2013}.

We further considered classical risk categories for product purchase by Jacoby and Kaplan \cite{jacoby1972}, which have often been used to measure perceived risk in marketing research~\cite{conchar2004,kushwaha2013}. Additionally, we adapt risk categories by Featherman and Pavlou \cite{DBLP:journals/ijmms/FeathermanP03}, who already adapted Jacoby and Kaplan's categories for e-commerce settings.

We split the perceived security risk in four risk categories: `general'\footnote{Jacoby and Kaplan \cite{jacoby1972} denote this risk category as `overall'. We renamed it to `general' since `overall' could be misunderstood as average score over all risk categories.}, `privacy', `physical', and `financial'.
Jacoby and Kaplan \cite{jacoby1972} and Featherman and Pavlou \cite{DBLP:journals/ijmms/FeathermanP03} present further risk categories that we did not consider because they have low relevance for the security risk of IoT products: `performance', `time', `psychological', and `social'.
Although the performance of IoT products can be affected in a security incident, performance deficiencies that affect functionality in a dangerous way are already covered by physical risk.
The risk of wasting time in case of a security incident exists for all product categories alike.
We excluded psychological risk as its original definition relates to the consumer's self-image or self-concept regarding a product \cite{jacoby1972}\footnote{Definition of psychologocal risk~\cite{jacoby1972}: ``the chances that an unfamiliar brand of [product] will not fit in well with your self-image or self-concept.''}.
Effects of IoT products on consumers' psychological state (e.g., perception of surveillance, or privacy violations) are considered in our scale by items in the risk categories `privacy' and `general'.
Finally, we did not take social risk into account since privacy risk already covers effects on the status in one's social groups.

Item candidates were generated and iteratively improved through expert reviews by 14 experts from the domains of cybersecurity, psychology and marketing research. % Norman, Werner, Johannes, Marcel, Tobias, Gaston, (Zina), (Philipp), (CM), (Felix), Feedback von 4 Kollegen von Christoph 
The final scale is presented in Appendix, \Cref{tab:scale-items} and consists of 13 items relating to risk categories `general', `privacy', `physical', and `financial'.
For the statistical comparison of the product categories, we averaged the scale to form a composite index.

\subsubsection*{Survey Structure and Data Collection}
\label{sec:study:prestudy:data}

We selected eight candidate product categories through an overview of popular IoT consumer products on online shopping websites and expert judgment: smart alarm systems, smart door locks, smart light bulbs, smart home cameras, smart smoke detectors, smart thermostats, smart vacuum robots, and smart weather stations.
In the surveys, the products were introduced in a randomized order.
Each product category was introduced with an exemplary product picture and a short text that explained the products' features and usage scenarios.  
We emphasized that all these products connect to the Internet.

To determine the sample size for this prestudy, we performed a power analysis \cite{faul2007} for paired $t$-tests. Assuming that large effects indicate practical relevance (Cohen’s $d_z$ = 0.8), and the desired power of 0.99, the power analysis determined 30 participants as sufficient.

We collected data with an online questionnaire using LimeSurvey \cite{limesurvey}.
The questionnaire was pretested by six experienced colleagues at our institutes. % Testers: Lena + Norman + 4 CM Kollegen
During the tests we realized that the amount of data that we wanted to collect would lead to a long and exhausting survey.
Therefore, we decided to split the survey into two smaller questionnaires that should be answered by two independent groups of respondents.
One group answered the perceived security risk (C1) for all eight product categories, while the other group evaluated the scales of C2 for all eight product categories.
Each group consisted of 30 crowdworkers. Through test runs, we estimated the average time to answer the surveys to be 10 to 12 minutes. We paid each crowdworker \euro{}1.80 for 12 minutes. % and paid the crowdworkers accordingly

%Thus, we paid each crowdworker a reward of \euro{}1.80 (reward equivalent for 12 minutes), adding up to overall cost of \euro{}3.00 per respondent including platform fees and taxes. 

\subsubsection*{Results}
\label{sec:study:prestudy:results}

Sixty respondents (23 female, 37 male) aged between 19 and 62 years ($\mu$ = 38.5, $\sigma$ = 10.7) answered the surveys.
We did not exclude any responses.
The collected data was analyzed using IBM SPSS \cite{field2013-spss}. %, a widely used software for statistical analysis in social science.  
The perceived security risk scale (C1) showed good statistical properties, which are not presented here for brevity. 
However, in \Cref{app:details:pca}, we present the statistical properties of the scale using the results of the main study (\Cref{sec:study:conjoint}).
For all scales of C2, Cronbach's alpha, a measure that defines the inner consistency of a scale, was above the recommended threshold of .700 (\textgreater .858)  \cite{field2013-spss}. 

%We evaluated all scales of C2 in terms of the Cronbach's alpha, a measure that defines the inner consistency of a scale. For all factors, Cronbach's alpha was above the recommended threshold of .700 (\textgreater .858)  \cite{field2013-spss}. 

%The product category smart alarm system was associated with the highest security risk, while smart weather stations had the lowest perceived security risk.

According to our criteria, we found three candidate pairs of product categories that do not statistically significantly differ from each other in the factors of C2, but  differ statistically significantly in the perceived security risk:

\begin{enumerate}
\item Smart home camera and smart weather station \\($t(29) = 7.57, p < 0.001, d_z = 1.383$)
\item Smart smoke detector and smart thermostat \\($t(29)= 2.09, p<0.05, d_z = 0.381$)
\item Smart smoke detector and smart vacuum robot \\($t(29) = 3.29, p < 0.01, d_z = 0.600$)
\end{enumerate}

We decided on the first pair as these product categories have the highest difference between their perceived security risk scores.
More analysis details can be found in  Appendix, Tables~\ref{tab:app:significance} and~\ref{tab:app:results-experiment1}.

%%%%%%%%%%%%%%%%%%%%%%%%%%%%
% PRODUCT ATTRIBUTES
%%%%%%%%%%%%%%%%%%%%%%%%%%%%

\subsection{Prestudy 2: Definition of Product Attributes and Levels} 
\label{sec:prestudy:attributes}

After two product categories were chosen, the next step was to determine product attributes that will be used in the conjoint analysis.
The number of attributes should be reasonable such that a respondent can process them cognitively~\cite{green1978}.
Otherwise respondents might tend to use shortcut heuristics that ignore less important features~\cite{eggers2009}.
%Prior research \philipp{Find better ref:} \cite{steiner-meissner} shows that if more than 6 to 8 attributes are used in product profiles, the respondents tend to use shortcut heuristics that ignore less important features.
%Prior research \cite{eggers2009} states that the number of attributes used in product profiles should kept to a realistic amount, otherwise respondents might tend to use shortcut heuristics that ignore less important features.
We decided for 7 attributes per product category. 
Two attributes were reserved for the attributes of the \seclabel{} (availability period and provisioning time).
The remaining five attributes comprised existing product attributes that depend on the product category. 
We paid attention to avoid correlation between attributes, which would lead to illogical profiles.
In the literature on conjoint analysis, the specification of attributes lacks a golden standard \cite{louviere2010} and is approached in various ways, such as focus groups, surveys, or expert judgements.

\subsubsection*{Method}
We conducted an online survey to identify the most important attributes for each product category to use in conjoint analysis. 
For this, attribute candidates were collected from online shopping websites. 
We prepared an online questionnaire on LimeSurvey that listed these 18 individual attributes, which are given in Appendix, \Cref{tab:app:attributes-scores}.
Respondents rated the importance of these attributes for their buying decision using the dual-questioning methodology by Alpert \cite{alpert-1971,voelckner2008}.
The respondents rated following two items on 7-point Likert scales:
``How important is each of these attributes in your buying decision?'' and
``How much difference do you feel, there is among products of the product category `[product]' in each of these attributes?''.
Both scores were multiplied to get an overall score for each product attribute.
The higher the overall score, the more important is the attribute.

\subsubsection*{Data Collection}
We recruited 30 crowdworkers for this preliminary survey who were each paid \euro{}0.75 (reward equivalent for 5 minutes).
In the screening of the collected data, we found that one participant clicked the middle option for almost all attributes and was six times faster than the average participant.
This person was excluded from the evaluation~\cite{barge2012}.

\subsubsection*{Results}
The final set of participants consisted of 29 respondents (18 female, 11 male) in the age between 20 and 63 years ($\mu$ = 35.3, $\sigma$ = 11.5).
The results are reported in Appendix, \Cref{tab:app:attributes-scores}.
In summary, the product attributes that are perceived as most important for smart home cameras were price, resolution, field of vision, frame rate, and zoom function.
For smart weather stations, the most important product attributes were price, battery lifetime, precision, rain and wind sensor, and expandability for multiple rooms.
In Appendix, \Cref{tab:app:attributes-scores}, we listed `solar panel for energy generation' as second most important product attribute for smart weather stations.
However, in line with prior research \cite{holbrook1990}, we refrained from this attribute because it correlates with `battery lifetime', and the correlated attributes bear the risk of threatening the study's validity.
We decided to include `battery lifetime' as nearly all smart weather stations use batteries, while solar panels are only a rare feature in this product category.
In addition to these attributes,  we added the availability and provisioning time of security updates introduced in \Cref{sec:label:idea}.
The final list of attributes is shown in \Cref{tab:attributes}. 

\begin{table}
  \centering
  \captionof{table}{Product categories with their respective attributes and  attribute levels.}
  \label{tab:attributes}
      \renewcommand{\arraystretch}{1}
  \begin{tabular}{@{}L{0.9cm} L{0.0cm} L{2.0cm} L{3.8cm}}
    \toprule
    \bf Category & \multicolumn{2}{l}{\bf Attribute} & \bf Levels \\
    \midrule
    \multirow{9}{0.9cm}{Smart home camera} &
    1 & \multicolumn{1}{L{2.0cm} }{Price} & \euro{}100, \euro{}120, \euro{}140, \euro{}160\\ 
%    \cmidrule{3-4}
    \multicolumn{1}{ l }{}                        &
    2 &\multicolumn{1}{ L{2.0cm}  }{Resolution} & HD, Full-HD  \\ 
%    \cmidrule{3-4}
    \multicolumn{1}{ l }{}                        &
    3 & \multicolumn{1}{L{2.0cm}  }{Field of vision} & 110\textdegree, 130\textdegree, 150\textdegree \\ 
%    \cmidrule{3-4}
    \multicolumn{1}{ l }{}                        &
    4 & \multicolumn{1}{ L{2.0cm} }{Frame rate} & 25fps, 30fps, 50fps \\  
%    \cmidrule{3-4}
    \multicolumn{1}{ l }{}                        &
    5 & \multicolumn{1}{L{2.0cm}  }{Zoom function} & yes, no  \\ 
%    \cmidrule{3-4}
    \multicolumn{1}{ l }{}                        &
    6 & \multicolumn{1}{L{2.0cm}  }{Availability of security updates} & none,  until 12/2020 (2 years), until 12/2024 (6 years)  \\ 
%    \cmidrule{3-4}
    \multicolumn{1}{ l }{}                        &
    7 &  \multicolumn{1}{L{2.1cm}  }{Provisioning time for  sec.\ updates} & none, within 10 days, within 30 days  \\ 
    \midrule
    \multicolumn{1}{@{}L{1.0cm}  }{\multirow{9}{1.0cm}{Smart weather station} } &
    1 & \multicolumn{1}{ L{2.0cm} }{Price} & \euro{}100, \euro{}120, \euro{}140, \euro{}160\\  
%    \cmidrule{3-4}
    \multicolumn{1}{ l }{}                        &
    2 & \multicolumn{1}{ L{2.0cm}  }{Battery lifetime} & 1 year, 2 years, 3 years  \\ 
%    \cmidrule{3-4}
    \multicolumn{1}{ l }{}                        &
    3 & \multicolumn{1}{ L{2.0cm}  }{Precision} &  $\pm$0.2\textdegree{}C, $\pm$0.3\textdegree{}C, $\pm$0.5\textdegree{}C  \\ 
%    \cmidrule{3-4}
    \multicolumn{1}{ l }{}                        &
    4 & \multicolumn{1}{L{2.0cm}  }{Rain/wind sensor} & yes, no \\  
%    \cmidrule{3-4}
    \multicolumn{1}{ l }{}                        &
    5 & \multicolumn{1}{L{2.0cm}  }{Expandability to multiple rooms} & yes, no  \\ 
%    \cmidrule{3-4}
    \multicolumn{1}{ l }{}                        &
    6 & \multicolumn{1}{L{2.0cm}  }{Availability of security updates} & none,  until 12/2020 (2 years), until 12/2024 (6 years)  \\ 
%    \cmidrule{3-4}
    \multicolumn{1}{ l }{}                        &
   7 &  \multicolumn{1}{L{2.1cm}  }{Provisioning time for  sec.\ updates} & none, within 10 days, within 30 days  \\ 
    \bottomrule
  \end{tabular}
\end{table}

\subsubsection*{Attribute Levels}
For each attribute, a discrete number of levels had to be specified.
The number of attribute levels should be as low as possible and does not need to cover the full feature range of the attribute. 
As recommended \cite{scholl2005}, we defined two to four levels to keep the complexity of the conjoint analysis low.
Similarly to the specification of the attributes, there is no golden standard in defining the attribute levels.
We decided to specify the levels with reasonable values that we acquired through the examination of the most popular products in both product categories available on Amazon as of October 2018. 
Thereby, we took care to avoid the specification of extreme values that are considered outliers. % \cite{bridges2011}.  

%% Price
We considered price levels in the realistic price ranges on the market.
We found smart home cameras by 13 manufacturers with prices between \euro{}50 and \euro{}179 (around \euro{}110 on average), and smart weather stations by 10 manufacturers in the range between \euro{}54 and \euro{}176 (around \euro{}129 on average).
To enhance the comparability, we defined the same price levels for both product categories. 
Furthermore, we decided not to use prices below \euro{}100 as the threshold between a two-figure price and a three-figure price might bias the importance of the price attribute considerably towards two-figure prices. % German: \cite{diller1996,muller1983}. 
Also, we consistently used 0-ending prices \cite{manoj2005, baumgartner2007} and finally chose four price levels with an equal spacing  between \euro{}100 and \euro{}160.
For all further product attributes, we chose two to three levels that reflect the attribute span of products on the market.
All attribute levels are summarized in \Cref{tab:attributes}.

%% SLL Focus Group
\subsubsection*{Focus Group}
As the attributes of the \seclabel{} are unknown to consumers, we run a focus group to gain an intuitive description of the \seclabel{}'s attributes and their levels.
%\footnote{Prior to this session, we referred to the label as `security lifetime label' with the attributes `security lifetime' and `time-to-patch'. % \cite{DBLP:conf/wisec/MorgnerFB18}.
%Due to comments, we realized that `security lifetime' might convey a false sense of security: These labels could be conceived as a guarantee that the security of the product is ensured for its full lifetime.
%Thus, we changed the name to `\seclabel{}s' and the renamed the attributes as `availability period' and `provisioning time'.}.  %, we run a focus group. % to determine the best way to intuitively describe these attributes with few words.
The focus group consisted of 8 participants (5 females and 3 males) in the age between 19 and 54 years ($\mu$ = 33.5, $\sigma$ = 12.9) without professional cybersecurity background.
Two participants had a professional IT background.
Thus, we paid attention that these two participants did not dominate the discussion.
We rewarded each participant with \euro{}10 for a one hour session.
We started the focus group by establishing the participants' prior experience with IoT consumer products, their awareness of security problems in these products, as well as their experience with security updates in general.
Then, we introduced the idea of the \seclabel{}s and asked the participants to write down how they would explain the attributes to family and friends.
Each participant presented their explanations and we discussed them with the group.
%During the focus group, we also asked participants to estimate the hypothetical lifetime of a smart home camera and smart weather stations in their homes.
%The results showed that in average the lifetime of a smart weather station with around 8-9 years is estimated  longer than the lifetime of a smart home camera with around 5-6 years.
%\zina{why was this asked, did it have influence on the attribute levels?}
%Furthermore, we asked for the provisioning time that they would wish for these products as well as what provisioning time they think is realistic.
%Interestingly, the ideal provisioning time for both product categories would be between 0 and 3 days, while the participants estimated that a realistic time-to-patch would be between 2 and 28 days. 
%In the last phase of the focus group, the participants reviewed the proposed product attribute levels for the conjoint analysis and provided feedback for clarification.  

Based on the focus group discussion, we presented the availability period in the conjoint analysis survey as `availability of security updates' (German: `Verfügbarkeit von Sicherheitsupdates') with a fixed end date, e.g., `until 12/2020', and with the relative period of time until this date (e.g. `2 years') to reduce the cognitive effort for the respondents.
Although the relative period is not part of the proposed label, the reduction of cognitive load was important as the respondents will compare availability attributes in 10 choice tasks.
The levels of the availability attribute were chosen to reflect realistic conditions: There is always a level of non-availability (i.e., manufacturer does not guarantee security updates), a level similar to the usual warranty period of this class of products (i.e., 2 years), and a third level that exceeds the usual warranty period and is more oriented on the realistic lifetime of the product (i.e., 6 years).
The provisioning time attribute was defined to reflect non-availability, a rather fast (and ideal) period of 10 days, and a slower (and more realistic) period of 30 days.
We denoted the provisioning time attribute in the survey as `provision time of security updates' (German: `Bereitstellungszeit von Sicherheitsupdates').

To exclude confusing profiles that occur from certain combinations of availability and provisioning time, e.g., the manufacturer does not provide security updates but offers a provisioning time of 30 days, we only allowed for meaningful combinations of both attributes.

%%%%%%%%%%%%%%%%%%%%%%%%%%%%
% CONJOINT
%%%%%%%%%%%%%%%%%%%%%%%%%%%%

%%%%%%%%%%%%%%%%%%%%%%%%%%%%
% CONJOINT
%%%%%%%%%%%%%%%%%%%%%%%%%%%%
\section{Conjoint Analysis }
\label{sec:study:conjoint}

We decided on a choice-based conjoint (CBC) analysis as this variant is the de-facto conjoint data collection standard in marketing research \cite{sawtooth2017} (cf.\ \Cref{sec:background:conjoint}). 
Although CBC's data collection is considered less efficient than other conjoint data collection methods, it provides a better predictor of real-world in-market behavior \cite{pinnell2005}. 
Furthermore, it allows a ``no choice''-option that also contributes valuable information, i.e., that all options are unattractive.

\subsection{Method}
\label{sec:study:conjoint:survey}

We used Lighthouse Studio by Sawtooth Software \cite{sawtooth2017} for survey setup and data analysis.
Lighthouse Studio is a well-established and validated tool for conjoint analysis \cite{natter-feurstein-2002,miller-2011}. %\cite{hofstede2002} %sawtooth-references,
% https://www.sawtoothsoftware.com/products/conjoint-choice-analysis/conjoint-analysis-software

To avoid fatigue, each respondent evaluated only one of the two product categories, whereas the product category was assigned randomly to the respondents.
Upon starting the survey, general information about the context, privacy of collected data, and the scope of the survey were presented. 
The respondents expected a survey about a smart home product. 
Then, the particular product category was introduced with a short explanation about its features and exemplary product pictures. 
We asked if the respondent is familiar with this product category and if she owns such a product.

In line with previous research \cite{huber1993,steenkamp1994}, we explained all attributes shown in \Cref{tab:attributes} (except price) with a short description, to raise their comprehension and  prevent misunderstandings.
To validate the comprehension, the respondent answered a quiz that included a question for each attribute.
For example, for the availability attribute, we asked ``What does the availability of security updates specify?'' with possible answers: (a) ``for how long the manufacturer guarantees to provide security updates'', (b) ``for how long the device is allowed to be used'', (c) ``for how long the device guarantees to be protected against hacker attacks''.
While the first answer was correct in this example, the order of possible answers was randomly permuted in the questionnaire. 
If the respondent chose a wrong answer, the correct answer was explained again.
Also, the quiz did not follow the order of how the attributes were presented before to rule out learning effects.

%\philipp{(10) How is the label presented?}
After the respondent became familiar with the attributes, the choice tasks for the conjoint analysis were explained: 
The respondent is presented with four product profiles and has to decide for the most attractive option.
All product profiles are described by an attribute level for each product attribute (including the \seclabel{} attributes) in plain text.
In addition to the set of product alternatives, there is always a ``no choice''-option that can be chosen in case none of the four profiles is desirable.
An exemplary choice task is depicted in Appendix, \Cref{fig:choice-task}.
Then, each respondent performed ten choice tasks including eight randomly-generated tasks, which were individual for each respondent, and two fixed holdout tasks, which were identical for all.
The fixed holdout tasks were later used to validate the attribute preference model. This model is based on the eight randomly generated tasks and should be similar to the model based on the holdout tasks in order to ensure internal consistency of the conjoint analysis (\Cref{sec:study:conjoint:validity}).  %, which in turn is solely based on the randomly-generated tasks
The respondents were not aware whether a choice task is randomly generated or fixed.
Prior research~\cite{steiner2016} concluded that the order of the attributes affects the choice behavior.
Therefore, we randomly permuted the order of all 7 attributes for each respondent but kept the same order for a particular respondent as it might be confusing otherwise.

After the respondents performed the choice tasks, we measured the perceived security risk for the particular product category by facilitating the perceived security risk scale from \Cref{sec:prestudy:categories}. 
Finally, we used psychometric scales to measure the respondents' privacy concerns with the Internet \cite{dinev-hart-2006} as well as their security behavior intentions \cite{DBLP:conf/chi/SawayaSCKNY17}. 
%For all scales, we randomly permuted the order of items.

Next, we included control questions to assess whether the respondents had difficulties in understanding the survey, had been distracted, and if they took the choice tasks seriously.
These questions were used to identify unmotivated respondents that we later excluded from the analysis.
Finally, we collected demographic data: gender, year of birth, vocational qualification, professional IT background, and net income. 
This data was used to determine the representativeness of the sample as well as for the segmentation.
To ensure that the respondents stayed focused, we included a number of motivational statements in the questionnaire.

\subsection{Pilot Study}
\label{sec:study:conjoint:pilot}

The questionnaire was developed in multiple iterative rounds. 
After completing the final draft, we collected feedback from seven experts from academic and market research institutes. % Experten: Lena, Gaston, Melissa, 4 Kollegen von Christoph
We asked whether they understood the attributes and tasks, and if anything could be misleading. 
Using their feedback, we re-worded some instructions.
Finally, we tested the questionnaire with 60 crowdworkers to check that the questionnaire is working as expected, and to calculate the average task completion time needed to determine the compensation.

%The evaluation of the pilot study showed that it took around 8 minutes to complete the survey on average.

%Their feedback contained helpful advises, e.g., concerning the verbalization of instructions

\subsection{Sample Size}
\label{sec:study:conjoint:data}

%The preparation of the data collection consisted of the determination of the required sample size and the recruitment of respondents.
%
The sample size, i.e., number of respondents for our questionnaire, was chosen as a trade-off between increasing costs and decreasing sampling errors.
Sampling error arises if the samples of respondents do not represent the population. 
Practical guidelines \cite{orme2010} on CBC analyses recommend at least 300 respondents for studies without segmentation.
If a segmentation analysis is desired, as is the case with our study, then a minimum of 200 respondents per subgroup is advised. 
Since we aimed for a comparison of up to three subgroups, which is a usual configuration in a segmentation analysis, we decided to recruit around 800 respondents for each product category, and thus, 1,600 respondents in total.

In the pilot study, the average time to answer the questionnaire was 8 minutes.
Thus, we paid crowdworkers  \euro{}1.20.
%, adding up to overall cost of \euro{}2.00 including platform fees and taxes. 
We ensured that respondents of the prestudies could not participate in the main survey.

\subsection{Sample Characteristics}
\label{sec:study:conjoint:sample}

% Table: Demographics
\begin{table}
  \centering
  \captionof{table}{Demographic data of conjoint analysis sample compared with the German population. }
  \label{tab:demographics}
  \setlength{\tabcolsep}{9.2pt}
  \renewcommand{\arraystretch}{0.8}
  \begin{tabular}{@{}l l r r r@{}}
    \toprule
    & & \multicolumn{2}{l}{\bf Sample} & \bf Population\\
    \midrule
    	All &  & 1,466 &  & \\
    \midrule
	\multirow{3}{*}{Gender} 
	& female & 640 & 44.0\% & 49.8\%$^a$ \\ 
	& male & 805 & 55.3\% & 50.2\%$^a$ \\ 
	& 3rd option & 10 & 0.7\% &  \\     
     \midrule
	\multirow{4}{*}{Age (in years)} 
	& 18-24 & 339 & 23.1\% & 12.9\%$^a$ \\ 
	& 25-29 & 288 & 19.6\% & 9.6\%$^a$ \\ 
	& 30-49 & 432 & 29.5\% & 18.8\%$^a$ \\ 
	& 50-64 & 404 & 27.5\% & 58.7\%$^a$ \\ 
%    \midrule
% 	\multirow{4}{*}{Education} 
%	& none & XXX & YY.YY\% & ZZ.ZZ\% \\ 
%	& low\footnote{Includes German education level `Lehre' (apprenticeship).} & XXX & YY.YY\% & ZZ.ZZ\% \\  % Lehre
%	& medium\footnote{Includes German education levels `Fachschulabschluss' and `Fachakademie/Berufsakademie' (both low ranking college degrees).}  & XXX & YY.YY\% & ZZ.ZZ\% \\  % Fachschulabschluss, Fachakademie/Berufsakademie
%	& high\footnote{Includes German education levels `Fachhochschule' (high ranking college degree), `Hochschule' (university degree), and `Promotion' (PhD).}  & XXX & YY.YY\% & ZZ.ZZ\% \\ % FH, Hochschule, Promotion
%%	& n/a & XXX & YY.YY\% & \\ 
%    \midrule
% 	\multirow{4}{*}{Education} 
%	& none 				& 156 & 10.6\% & 22.8\%* \\ 
%	& Lehre				& 330 & 22.5\% & 47.1\%* \\ 
%	& Fachschule			& 181 & 12.3\% & 11.3\%* \\ 
%	& Fachakademie 		& 71 & 4.8\% & 1.8\%* \\ 
%	& FH				& 228 & 15.6\% & 6.4\%* \\ 
%	& Hochschule 		& 477 & 32.5\% & 9.3\%* \\ 
%	& PhD 				& 23 & 1.6\% & 1.4\%* \\ 
    \midrule    
 	\multirow{3}{1.7cm}{Vocational qualification} 
	& none 				& 156 & 10.6\% & 22.8\%$^a$ \\ 
	& vocational 			& 582 & 39.7\% & 60.2\%$^a$ \\ 
	& academic			& 728 & 49.7\% & 17.0\%$^a$ \\ 
    \midrule
 	\multirow{5}{1.7cm}{Monthly net income (in \euro{})} 
	& none 					& 52 & 4.3\% & 17.9\%$^b$ \\ 
	& less than 900 		& 288 & 23.5\% & 24.3\%$^b$ \\ 
	& 900 to 1,500 & 296 & 24.2\% & 22.7\%$^b$ \\ 
	& 1,500 to 2,600 & 377 & 30.8\% & 22.4\%$^b$ \\ 
	& more than 2,600 		& 210 & 17.2\% & 10.2\%$^b$ \\ 
%	& n/a & XXX & YY.YY\% &  \\ 
    \midrule
 	\multirow{2}{1.7cm}{Professional IT background} 
	& yes & 249 & 17.6\% & \multirow{2}{*}{~}  \\ 
	& no & 1166 & 82.4\% &  \\ 
%%	& n/a & XXX & YY.YY\% &  \\ 
    \bottomrule
  \end{tabular}
   \begin{tablenotes}
      \footnotesize
      \item $^a$Census 2011 (age 18-65)  \cite{german-census2011}, ~~$^b$Mikrozensus 2014 (all ages) \cite{german-mikrozensus2014} 
      \item Missing answers are ignored for percent proportioning.
  \end{tablenotes}
\end{table}

% Eventuell später: https://tex.stackexchange.com/questions/211898/putting-several-footnotes-below-the-table?rq=1

We collected the data within a week in mid-December 2018. % 3 days
After a screening, we excluded 154 (9.5\%) of the 1,620 collected data sets.
We excluded 70 data sets due to low task completion times (within less than half of the pilot study's average time), 48 data sets due to indications in the control questions, 19 data sets due to suspected multi-participation (same IP address and user agent), 16 data sets that answered more than two quiz questions wrong, and 3 data sets of respondents under 18 years.  
The final sample included 1466 data sets (640 female, 805 male) in the age between 18 and 65 years ($\mu$ = 33.8, $\sigma$ = 11.2).
Details of the demographic data are presented in \Cref{tab:demographics}.
In comparison to the German population, the sample is biased towards males and high-educated persons. 
Furthermore, people in the age of 50 and above are underrepresented in this sample.
However, the sample aligns to the target group of consumers interested in  IoT consumer products, which is likewise biased towards males, age group 25-34, and higher incomes \cite{statista-report2019}.

\subsection{Results}
\label{sec:study:conjoint:results}

%From the sample, 

In total, 731 respondents evaluated the product category `smart home camera' and 735 respondents assessed the product category `smart weather station'.
The high difference in the security risk perception between both product categories was confirmed:
While the smart weather station achieved an average perceived security risk score of 3.65, the smart home camera achieved an average score of 5.50. This difference is highly statistically significant with the large effect size: $t(1464) = 28.42, p < 0.001, d = 1.48$. Cronbach's alpha for all psychometric scales was above the recommended threshold of .700 (\textgreater .837) \cite{field2013-spss}.
%This aligns with results of the prestudy (cf. \Cref{sec:prestudy:categories}). 

%We evaluated the psychometric scales of the conjoint questionnaire in terms of Cronbach's alpha. 

%We evaluated the psychometric scales of the conjoint questionnaire in terms of Cronbach's alpha. 
%For all factors, Cronbach's alpha was above the recommended threshold of .700 (\textgreater .837) \cite{field2013-spss}. 

% Relative Importance
For the analysis of the collected conjoint data  from the randomly-generated choice tasks, we used hierarchical Bayes estimation with default settings as recommended  \cite{sawtooth2017}.
Using this estimation method, we determined the average relative importance of each product attribute as well as the part-worth utility for each product level based on a total of 5,848 (home camera) and 5,880 (weather station) choice tasks.
The relative importance of an attribute defines the relative impact (measured in percent) on the overall choice.
The importances are ratio data meaning that an attribute with an importance of 20\% is twice as much important as an attribute with an importance of 10\% \cite{orme2002}.
%For instance, a relative importance of 10\% means that this particular attribute has an impact of 10\% on the consumer's buying decision, while the other attributes account for 90\% of the decision. 
%The results are presented in \Cref{tab:rel-importance} and \Cref{tab:utilities}.

% Table: Rel. Importances + Utilities
\begin{table}[t]
  \centering
  \captionof{table}{Relative importance of product attributes.}
  \label{tab:rel-importance}
  \setlength{\tabcolsep}{1.5pt}
  \renewcommand{\arraystretch}{1}
%  \begin{minipage}[t]{0.23\textwidth}
  \begin{tabular}{@{}c l r r@{\hspace{3ex}}l r r@{}}
  \toprule
    & \multicolumn{3}{c}{\textbf{Smart Home Camera} (n=731)} & \multicolumn{3}{c}{\textbf{Smart Weather Station} (n=735)}\\
    \cmidrule(rl){2-4}
    \cmidrule(){5-7}
    \it Rank & \it Attribute & $\mu$ [\%] & $\sigma$ [\%] & \it Attribute & $\mu$ [\%] & $\sigma$ [\%]\\
    \cmidrule(){1-1}
    \cmidrule(rl){2-4}
    \cmidrule(){5-7}
    1. &	Availability\textsuperscript{1} & \multirow{1}{*}{30.57} & \multirow{1}{*}{9.82}   & Availability\textsuperscript{1}  & \multirow{1}{*}{20.37}  & \multirow{1}{*}{9.28}  \\
    2.	&	Price & 15.12 & 8.86	& Price & 16.64  & 11.03  \\
    3.	&	Provisioning time\textsuperscript{2} & \multirow{1}{*}{13.98} & \multirow{1}{*}{6.30} & Rain/wind sensor & 16.37  & 9.87  \\ 
    4.	&	Resolution & 12.05 & 8.98 		& Provisioning time\textsuperscript{2} & \multirow{1}{*}{16.12}  & \multirow{1}{*}{7.96}  \\
    5.	&	Frame rate & 10.71 & 6.06  		& Expandability & 13.15  & 8.14  \\
    6.	&	Field of view & 8.89 & 5.29  		& Battery lifetime & 9.82  & 6.53  \\
    7.	&	Zoom function & 8.68 & 6.72  		& Precision & 7.53  & 4.84  \\
    \bottomrule
  \end{tabular}
%  \end{minipage}
%  \begin{minipage}[t]{0.24\textwidth}
%  \begin{tabular}{@{}l r r@{}}
%  \toprule
%\multicolumn{3}{c}{\bf Smart Weather Station (n=735)}\\
%    \midrule
%     \it Attribute & $\mu$ [\%] & $\sigma$ [\%]\\
%    \midrule
%    	 & Availability  & \multirow{1}{*}{20.37}  & \multirow{1}{*}{9.28}  \\
%    	 & Price & 16.64  & 11.03  \\
%       & Rain/wind sensor & 16.37  & 9.87  \\ 
%    	& Provisioning time & \multirow{1}{*}{16.12}  & \multirow{1}{*}{7.96}  \\
%    	& Expandability & 13.15  & 8.14  \\
%    	& Battery lifetime & 9.82  & 6.53  \\
%    	& Precision & 7.53  & 4.84  \\
%    \bottomrule
%  \end{tabular}
%  \end{minipage}
   \begin{tablenotes}
      \footnotesize
      \item \textsuperscript{1}Availability of security updates, \textsuperscript{2}Provisioning time for security updates
  \end{tablenotes}
  \end{table}

\begin{table}[t]
  \centering
  \captionof{table}{Average utilities of selected product attributes.} %(zero-sum)\textsuperscript{1}
  \label{tab:utilities}
  \setlength{\tabcolsep}{4.4pt}
    \renewcommand{\arraystretch}{0.9}
  \begin{tabular}{@{}l l r r@{}}
  \toprule
  \bf \multirow{2}{*}{Attribute} & \bf \multirow{2}{*}{Level} & \multicolumn{1}{R{1.6cm}}{\bf Smart Home Camera} & \multicolumn{1}{R{1.9cm}@{}}{\bf Smart Weather Station}\\
   \midrule
    	\multirow{4}{*}{Price}
    	& 100\euro & 33.64 & 44.13  \\
    	& 120\euro & 19.46 & 21.08  \\
    	& 140\euro & -5.66 & -9.33 \\
    	& 160\euro & -47.45 & -55.88  \\
   \midrule
    	\multirow{3}{1.7cm}{Availability of sec.\ updates}
    	& none & -111.93 & -77.52  \\
    	& until 12/2020 (2 years) & 14.32 & 21.91  \\
    	& until 12/2024 (6 years) & 97.61 & 55.61  \\
   \midrule
    	\multirow{3}{1.7cm}{Prov.\ time for sec.\ updates}
    	& none & -32.82 & -61.10  \\
    	& within 10 days & 37.16 & 43.75  \\
    	& within 30 days & -4.34 & 17.35  \\
    \bottomrule
  \end{tabular}
  \begin{tablenotes}
      \footnotesize
			\item Utilities of all levels of a certain product attribute add up to zero.
%      \item Utilities of the remaining product attributes are shown in Appendix, \Cref{tab:utilities-appendix}.
  \end{tablenotes}
\end{table}

As listed in \Cref{tab:rel-importance}, the availability of security updates (31\%) is the most important attribute for  smart home cameras. It is twice as important as price (15\%) and provisioning time for security updates (14\%).
Other functional attributes are  less important with relative importance between 8\% and 12\%.
For the smart weather stations, the availability of security updates (20\%) is also the most important attribute.
However, the difference in importance to other product attributes is smaller than for the smart home cameras.
The relative importance of price, rain and wind sensor, and provisioning time for security updates are at around 16\%.
All other attributes are ranked with relative importances between 7\% and 13\%.

%\philipp{Revised:}
In general, the relative importance of the attributes differ between both product categories. Good research practice, e.g., \cite{baumgartner2007}, strongly discourages quantitative comparison of preference measurements of different product categories to each other, %should not be quantitatively compared 
as the importance of an attribute can only be interpreted as relative value \emph{within} the particular product category. Therefore, we discuss the possible differences qualitatively.
For the product with the high perceived security risk, especially the availability of security updates (twice as important as other attributes) seems to play a more crucial role in buying decisions than for the product with the low perceived security risk (only slightly more important than other attributes).
Furthermore, the provisioning time for security updates is considered the third most important attribute for the product with the high perceived security risk and the fourth most important attribute for the product with the low perceived security risk. 
Thus, the provisioning time for security updates seems to have a similar importance than price and other highly-ranked technical attributes.

To summarize, the availability of security updates plays the most important role in the consumers' choice in this study. The relative importance of the provisioning time for security updates is also high, although substantially lower than of the availability of security updates.
For both product categories, the second most important attribute is the price.
Functional features of both product categories are rated as less important than the attributes of the \seclabel{} and price, except the rain and wind sensor for smart weather stations.

\Cref{tab:utilities} shows the average utilities that consumers ascribe to the levels of product attributes \cite{orme2002}.
Negative utilities represent unfavorable options compared to the other options, while positive utilities describe the favorable options. The utilities of all levels of a certain product attribute add up to zero.
For the price attribute, lower prices have a higher utility for consumers.
For both product categories, the non-availability of security updates is considered as especially unfavorable with the highest negative utility scores among all attributes. The availability of security updates for 6 years is more favorable than  for 2 years.
The utility for the provisioning time for security updates shows a preference for short time period (10 days) rather than a longer time period (30 days). A provisioning time of 30 days has a negative utility for the smart home camera, in contrast to the smart weather station. Thus, participants dislike long provisioning times for a product with high perceived security risk.

\subsection{Validity}
\label{sec:study:conjoint:validity}

We tested the internal consistency of our results by comparing simulations based on the
preference measurement results with choice data from holdout tasks (cf. \Cref{sec:study:conjoint:survey}). 
Furthermore, we validated the  preference for long availability periods and short provisioning times.
The results of the holdout tasks were not included  into the calculation of the preference measurement results.

For the holdout tasks, we defined five product profiles -- P1, P2, P3a, P3b, P4 -- for each product category.
Profiles P3a and P3b differ  only in their \seclabel{} attributes: In P3a, there was no guarantee for the availability and provisioning time of security updates. In P3b, these two attributes were set to the objectively best levels: availability of security updates for 6 years with provisioning time of 10 days.

In the course of the survey, each respondent performed 10 choice tasks, of which 8 were randomly-generated tasks and 2 were fixed holdout tasks.
The fixed holdout tasks were the same for all respondents of a product category and appeared as the 4th and 8th of the 10 choice tasks. 
Both holdout task consisted of the four product profiles that were presented in a different fixed order.
In the first holdout task (i.e., the 4th choice task), the profiles (P1, P2, P3a, P4) were presented. In the second holdout task (i.e., the 8th choice task), the profiles (P4, P3b, P2, P1) were presented.

% Table: Holdout Tasks
\begin{table}[t]
  \centering
  \captionof{table}{Comparing the market estimation regarding shares of preferences of four pre-defined product configurations with real preferences in fixed holdout tasks. }
  \label{tab:holdout}
    \setlength{\tabcolsep}{1.5pt}
    \renewcommand{\arraystretch}{1}
  \begin{tabular}{@{}l r r r r r r r r@{}}
    \toprule
    & \multicolumn{4}{c}{\bf Smart Home Camera } & \multicolumn{4}{c@{}}{\bf Smart Weather Station } \\
    \cmidrule(l){2-5}
    \cmidrule(l){6-9}
     & \multicolumn{2}{c}{Holdout Task 1} & \multicolumn{2}{c}{Holdout Task 2} & \multicolumn{2}{c}{Holdout Task 1} & \multicolumn{2}{c@{}}{Holdout Task 2} \\ 
    \cmidrule(){2-3}
    \cmidrule(l){4-5}
    \cmidrule(l){6-7}
    \cmidrule(l){8-9}
     & \multicolumn{2}{c}{\it P1, P2, P3a, P4} & \multicolumn{2}{c}{\it P4, P3b, P2, P1} & \multicolumn{2}{c}{\it P1, P2, P3a, P4} & \multicolumn{2}{c@{}}{\it P4, P3b, P2, P1} \\ % w/ updates  w/o updates $^1$ $^2$
    \cmidrule(){2-3}
    \cmidrule(l){4-5}
    \cmidrule(l){6-7}
    \cmidrule(l){8-9}
    \tabrotate{\pbox{1.2cm}{ Product Profile\ }} 
     &  Estimated & Real  &  Estimated & Real & Estimated & Real  & Estimated & Real\\
%     &  Simulation & Real  &  Simulation & Real &  Simulation & Real  &  Simulation & Real\\
%    \midrule
    \cmidrule(r){1-1}
    \cmidrule(){2-3}
    \cmidrule(l){4-5}
    \cmidrule(l){6-7}
    \cmidrule(l){8-9}
    P1 &  1.5\% & 2.7\% 	& 1.4\% & 2.9\% 		& 1.7\% & 3.3\% 		& 1.5\% & 2.7\% \\
    P2 &  4.8\% & 11.4\% 	& 2.4\% & 5.7\% 		& 22.1\% & 21.0\% 	& 15.8\% & 16.5\% \\
    P3 &  3.6\% & 6.6\% 	& 58.8\% & 53.4\% 	& 5.3\% & 9.5\% 		& 36.5\% & 32.7\% \\
    P4 &  84.4\% & 72.8\% 	& 34.6\% & 35.6\% 	& 61.1\% & 57.1\% 	& 39.0\% & 42.0\% \\
    None &  5.7\% & 6.6\% 	& 2.8\% & 2.5\% 		& 9.7\% & 9.1\% 		& 7.1\% & 6.1\% \\
    \bottomrule
  \end{tabular}
     \begin{tablenotes}
      \footnotesize
      \item P3a and P3b differ only in the \seclabel{} attributes: P3a does not guarantee security updates, whereas P3b guarantees availability of security updates until 2024 (6 years) and a provisioning time within 10 days.      
  \end{tablenotes}
\end{table}

The results in \Cref{tab:holdout} (`Real'-columns) for smart home cameras show that in the first holdout task 6.6\% decided for P3a without guarantee for security updates, while in the second holdout task 53.4\% chose P3b with best guarantee for security updates.
The same effect can be observed for the smart weather station: in the first choice task, 9.5\% choose P3a, while 32.7\% decided for P3b.
This validates the results that we derived from the randomly-generated choice tasks: 
The (non-)availability of security updates is an important factor in buying decisions, while this effect seems to be higher for products with high perceived security risk.

We tested the consistency of the hierarchical Bayes estimation with the market simulator from Lighthouse Studio, which represents a standard procedure to assess the internal validity of CBC models \cite{natter-feurstein-2002}.
This market simulator estimates the shares of preferences for the product profiles P1 to P4 based on the preference measurement results of the conjoint analysis. 
We compared the estimated market shares with the evidence gathered through the holdout tasks (see \Cref{tab:holdout}).
For example, in the first holdout task of the smart home camera, the market simulator estimated a market share of 84.4\% for P4, while the evaluation of the real choices showed that 72.8\% of the respondents decided for P4. 
Based on the comparison of estimated and real choices, we conclude that although the simulator does not exactly match the real choices, the preference measurement results are robust in estimating the order and magnitude of the overall preferences, which indicates a high consistency of the results.

% Table: Segmentation
\begin{table*}
  \centering
  \captionof{table}{Consumer segmentation via latent-class analysis.}
  \label{tab:segmentation}
  \setlength{\tabcolsep}{3.2pt}
    \renewcommand{\arraystretch}{0.8}
  \begin{tabular}{l l r r l r r r l}
    \toprule
    \multirow{6}{*}{\bf } & &  \multicolumn{3}{c}{\bf Smart Home Camera}  &  \multicolumn{4}{c}{\bf {Smart Weather Station}} \\
    \cmidrule(l){3-5}
    \cmidrule(l){6-9}
    & \multicolumn{1}{r}{\it} & \it Group 1 & \it Group 2 &  & \it {Group 1} & \it {Group 2} &  \it {Group 3} & \\ 
    \cmidrule{1-2}
    \cmidrule(l){3-4}
    \cmidrule(l){6-8}
     \multicolumn{2}{r}{Total number of respondents $n$\textsuperscript{1}} & 488 & 243 & \it Group 1 vs Group 2 & {182} & {221} & {332} & \it {Group 1 vs Group 2 vs Group 3\textsuperscript{2}} \\ 
    \cmidrule{1-2}
    \cmidrule(l){3-4}
    \cmidrule(l){5-5}
    \cmidrule(l){6-8}
    \cmidrule(l){9-9}
	\multirow{2}{*}{Gender} 
%	& female 			& 42.6\% & 44.9\%  & \multirow{2}{*}{${\chi}^2(1) = 0.42, \phi = .090$} & {43.4\%}  & {50.2\%} & {40.1\%} & \multirow{2}{*}{{${\chi}^2(1) = 4.40, V = 0.08$}}\\ 
	& female 			& 42.6\% & 44.9\%  & \multirow{2}{*}{{${\chi}^2(1) = 0.42, V = 0.024$}} & {43.4\%}  & {50.2\%} & {40.1\%} & \multirow{2}{*}{{${\chi}^2(2) = 4.40, V = 0.08$}}\\ 
	& male  				& 55.9\%  & 53.1\% &  & {53.8\%}  & {49.8\%} & {58.7\%} & \\ 
    \cmidrule{1-2}
    \cmidrule(l){3-4}
    \cmidrule(l){5-5}
    \cmidrule(l){6-8}
    \cmidrule(l){9-9}
	\multirow{1}{*}{Age} 
	& $\mu$  [years]			& 33.0 & 33.9 & $t(729) = -1.10, d = -0.09$ & {36.9} & {34.8} & {32.8} & {$F(2, 732) = 8.00\textsuperscript{**}, r = 0.15$}  \\ 
    \cmidrule{1-2}
    \cmidrule(l){3-4}
    \cmidrule(l){5-5}
    \cmidrule(l){6-8}
    \cmidrule(l){9-9}
 	\multirow{3}{1.7cm}{Vocational qualification} 
	& none 				 & 9.2\%   & 11.9\%  & & {8.2\%}   & {11.8\%} &  {12.3\%} & \\ 
	& {vocational}	 & 38.7\%  & 42.8\% & {${\chi}^2(2) = 3.35, V = 0.068$} & {37.9\%}  & {43.0\%} & {37.7\%} & {${\chi}^2(4) = 68.31\textsuperscript{**}, V = 0.22$} \\ 
	& academic			 & 52.0\%  & 45.3\% & & {53.8\%}  & {45.2\%} &  {50.0\%} & \\ 
    \cmidrule{1-2}
    \cmidrule(l){3-4}
    \cmidrule(l){5-5}
    \cmidrule(l){6-8}
    \cmidrule(l){9-9}
 	\multirow{1}{2.7cm}{Monthly net income\textsuperscript{3}} 
	& $\mu$	 [\euro{}]			& {1704}   & {1706}  &   {$t(599) = -0.22, d = 0.002$} & {1854} & {1707} & {1532} &  {$F(2, 619) = 3.73\textsuperscript{*},  r = 0.11$} \\ 
    \cmidrule{1-2}
    \cmidrule(l){3-4}
    \cmidrule(l){5-5}
    \cmidrule(l){6-8}
    \cmidrule(l){9-9}
 	\multirow{2}{1.7cm}{Professional IT background} 
	& yes 				& 18.5\%  & 17.7\%  & \multirow{2}{*}{{${\chi}^2(1) = 0.06, V = 0.01$}} & {13.7\%}  & {17.2\%} & {17.8\%} & \multirow{2}{*}{{${\chi}^2(2) = 0.37, V = 0.02$}} \\ 
	& no  				& 81.5\%  & 82.3\%  & & {81.3\%}  & {81.4\%} &  {80.4\%} &\\ 
    \cmidrule{1-2}
    \cmidrule(l){3-4}
    \cmidrule(l){5-5}
    \cmidrule(l){6-8}
    \cmidrule(l){9-9}
	\multicolumn{2}{l}{Privacy concerns} 	& 4.48 & 4.59 &  $t(729) = -0.98, d = -0.08$ & {4.54} & {4.57} & {4.26} & {$F(2, 732) = 3.86\textsuperscript{*},  r = 0.10 $} \\ 
	\multicolumn{2}{l}{Security behavior intention} 	 & 3.59 & 3.56 &  $t(729) = 0.59, d = 0.05$ & {3.70} & {3.67} & {3.47} &  {$F(2, 732) = 16.74\textsuperscript{**}, r = 0.17$} \\ 
    \cmidrule{1-2}
    \cmidrule(l){3-4}
    \cmidrule(l){5-5}
    \cmidrule(l){6-8}
    \cmidrule(l){9-9}
     	 \multicolumn{2}{l}{Perceived security risk} 
				& 5.44 & 5.61 &  $t(729) = -2.13\textsuperscript{*}, d = -0.17$ & {3.88} & {3.96} & {3.32} & {$F(2, 732) = 10.31\textsuperscript{**}, r = 0.21$}\\ 
    \bottomrule
  \end{tabular}
   \begin{tablenotes}
      \footnotesize
      \item Notation: Statistically significant with \textsuperscript{*}$p<0.05$, \textsuperscript{**}$p<0.01$, 
      \textsuperscript{1}If missing values appear for specific variables, respective analyses rely on a diverging number of cases.
      \textsuperscript{2}Post-hoc tests are given in Appendix, \Cref{tab:app:posthocs}.
% Alternative von CM:
%      {\textsuperscript{3}Thirteen participants indicated a monthly net income of “more than \euro{}5000”. Following \cite{hanneman-2012}, for analyses, values were set to the lower class boundary and multiplied with the factor 1.5 in line with the German Mikrozensus \cite{lengerer-2010}.}
      \textsuperscript{3}Income data were processed following the methodology of the German sample census \cite{lengerer-2010}. %,hanneman-2012
  \end{tablenotes}
\end{table*}

\begin{table*}
  \centering
  \captionof{table}{Product attribute importance of consumer groups. }
  \label{tab:segmentation2}
  \setlength{\tabcolsep}{3pt}
    \renewcommand{\arraystretch}{1}
  \begin{tabular}{c l r l r@{\hspace{3ex}}l r l r l r}
    \toprule
    & \multicolumn{4}{c}{\bf Smart Home Camera} & \multicolumn{6}{c}{\bf {Smart Weather Station}}  \\
    %\cmidrule(){1-1}
    \cmidrule(lr){2-5}
    \cmidrule(){6-11}
    \it Rank & \multicolumn{2}{c}{\it Group 1 (n = 488)} & \multicolumn{2}{c}{\it Group 2 (n = 243)} & \multicolumn{2}{c}{\it {Group 1 (n = 182)}} & \multicolumn{2}{c}{\it {Group 2 (n = 221)}}  & \multicolumn{2}{c}{\it {Group 3 (n = 332)}}  \\ 
    \cmidrule(){1-1}
    \cmidrule(l){2-3}
    \cmidrule(lr){4-5}
    \cmidrule(){6-7}
    \cmidrule(l){8-9}
    \cmidrule(l){10-11}
	1. & Availability\textsuperscript{1}  		& 31.42\%  	& Availability\textsuperscript{1}  		& 25.26\%  	& {Availability\textsuperscript{1}} 		& {23.41\%} & {Availability\textsuperscript{1}} 		& {35.24\%}	& {Price} 								& {29.98\%}	\\		
	2. & Resolution  							& 16.47\% 	& Provisioning time\textsuperscript{2}  	& 19.03\%  	& {Rain/wind sensor  	}				& {23.14\%}	& {Provisioning time\textsuperscript{2}}  	& {24.62\%}	& {Rain/wind sensor}  					& {20.55\%}  	\\
	3. & Price 								& 11.82\%  	& Price 								& 16.66\%  	& {Provisioning time\textsuperscript{2}} & {19.38\%}	& {Expandability } 						& {12.95\%}	& {Battery Lifetime}  					& {13.46\%}  	\\
	4. & Frame rate  							& 10.68\%  	& Field of view 						& 11.64\%  	& {Expandability} 						& {15.90\%}   & {Battery Lifetime } 					& {8.12\%}	& {Expandability}  						& {12.88\%}  	\\
	5. & Zoom function  						& 10.48\%   	& Frame rate 						& 9.26\%   	& {Price }							& {8.49\%}	& {Rain/wind sensor}  					& {7.11\%}	& {Precision}		 					& {8.49\%}   	\\
	6. & Provisioning time\textsuperscript{2} 	& 10.22\%  	& Resolution 							& 9.17\%  	& {Battery Lifetime  }					& {5.26\%} 	& {Precision}		 					& {6.19\%} 	& {Availability\textsuperscript{1}} 		& {8.01\%} 	\\
	7. & Field of view  						& 8.91\%  	& Zoom function 						& 8.98\%   	& {Precision	}	 					& {4.44\%}	& {Price} 								& {5.76\%} 	& {Provisioning time\textsuperscript{2}}  	& {6.64\%} 	\\
    \bottomrule
  \end{tabular}
   \begin{tablenotes}
      \footnotesize
      \item \textsuperscript{1}Availability of security updates, \textsuperscript{2}Provisioning time for security updates
  \end{tablenotes}
\end{table*}

\subsection{Segmentation}
\label{sec:study:conjoint:segmentation}

We used the latent class segmentation module of Lighthouse Studio to assign respondents to groups that have similar preferences.
The module implements latent class analysis, a classification technique to find groups in multi-dimensional data. 
First, we needed a measure to decide in how many reasonable segments we split the respondents.
Following recommendations of Sawtooth Software \cite{sawtooth-latent-class}, we used the consistent Akaike's information criterion \cite{bozdogan1987} as measure.
Based on this criterion (see \Cref{app:details:akaike}), we decided to split the respondents of the product categories ``smart home camera'' and ``smart weather station'' into two and three segments, respectively.
The consumer segmentation for both product categories is shown in \Cref{tab:segmentation}. 
The differences in preferences between the segments are given in \Cref{tab:segmentation2}.

For the smart home camera, both segments differ statistically significantly only in the perceived security risk. 
For the first group with lower security risk perception towards smart home cameras, the availability of security updates (31\%) is the most important product attribute and twice as important as resolution (16\%) and price (12\%).
In comparison to the importance of technical features (9--11\%) other than the resolution, the availability of security updates is even three times more important.
For the second group with the higher security risk perception, the availability (25\%) and provisioning time (19\%) of security updates are the most important product attributes followed by price (17\%).
Compared to the importance of the technical features (9--12\%), availability and provisioning time are (almost) twice as important.

For the smart weather station, the segments differ statistically significantly with regard to age, vocational qualification, income, privacy concerns, security behavior intentions, and perceived security risk. 
The first group exhibits the highest average age, high vocational qualifications, the highest income, high privacy concerns, the highest security behavior intentions, as well as a high security risk perception towards smart weather stations.
For this group, the availability of security updates (23\%) as well as the rain and wind sensor (23\%) are the most important product attributes, followed by provisioning time (20\%) and expandability (16\%).
These attributes are 2 to 3 times as important as the price (8\%).

The second group is characterized by lower vocational qualification, the highest privacy concerns, and the highest perceived security risk.
For this group, the attributes of the \seclabel{} dominate the choice decision and account for 60\% of the overall importance.
The availability (35\%) and provisioning time (25\%) is three and two times as important, respectively, than technical features. 
For this group, the price (6\%) is the least important attribute for their choice.

The third group perceives the lowest risk for smart weather stations.
This group is the youngest on average, has the lowest income, as well as the lowest privacy concerns and security behavior intentions.
The most important product attribute is the price (30\%), followed by technical features, such as rain and wind sensor (21\%) and battery lifetime (13\%).
The availability (8\%) and provisioning time (7\%) are the least important attributes, i.e., the \seclabel{} plays only a minor role in the consumers' choice of this segment.

%%%%%%%%%%%%%%%%%%%%%%%%%%%%
% DISCUSSION
%%%%%%%%%%%%%%%%%%%%%%%%%%%%

\section{Discussion}
\label{sec:discussion}

We discuss research questions formulated in \Cref{sec:study}, and consider economic and policy implications of our results.

%RQ1
\textit{RQ1: Relative importance of the availability period and provisioning time.} 
% For IoT consumer products with low and high perceived security risk, t
The availability of security updates was the product attribute with the highest relative importance (up to twice as important as other high-ranked attributes) as well as with the widest span of average utility (\Cref{tab:rel-importance}). %
Provisioning time for security updates was evaluated as less important than their availability, but seems to be more important than most technical features.
Consumers also prefer short provisioning times, and even assign a negative utility (i.e., dislike) to the provisioning time of 30 days for the product with the high perceived security risk (\Cref{tab:utilities}).
The high importance of the attributes of the \seclabel{} is surprising, as users are not familiar with these attributes.
This might be due to the explicit mentioning of the non-availability of security updates that might have discouraged the users.
This effect can be seen in \Cref{tab:utilities} where the negative utility of  non-availability is greater (-111.93 for home cameras and -77.52 for weather stations) than the positive utility of availability of security updates for 6 years (97.61 and 55.61, respectively). % (cf. \Cref{tab:utilities}).
This indicates that consumers want to avoid the non-availability of security updates, and therefore, the mandatory nature of the \seclabel{} is very important. 
%\zina{I tried to make generalizations for all RQs, as this is expected in the discussion.}

% rather than favor a long availability period.

% to buy products without a security update guarantee

\textit{RQ2: Differences between products with a low and high perceived security risk.} 
We observed differences in the relative importance of the security update attributes. 
For the product with the high perceived security risk, the importance of availability is at least twice as high as the importance of other attributes.
In contrast, availability is only slightly more important than for other high-ranked attributes for the product with a low perceived security risk.
Furthermore, for the product with the high perceived security risk, the results of the holdout tasks (cf.\ \Cref{tab:holdout}, `Real'-columns) show an increase of preference for product profile P3a from 6.6\% (without guaranteed security update) to 53.4\% for profile P3b (with 6-year guarantee for security updates).
For the product with the low perceived security risk, there is a comparatively smaller increase from 9.5\% to 32.7\% in the preference for the analogous product profiles.
Also, if we sum up the importance of both label attributes, the \seclabel{} shows a relative importance of 45\% for the consumers' choice regarding the product with the high perceived security risk, compared to 36\% for the product with the low perceived security risk.

%Thus, \seclabel{}s may influence the consumers' choice more strongly for products with  high perceived security risk than for products with low perceived security risk. 

This finding supports the concern of moral hazard (\Cref{sec:label:concerns}): Users might consider \seclabel{}s primarily if they think that they could be personally affected by security incidents. In reality, however, products with seemingly low perceived security risk for the owner can be used for serious attacks. For example, Mirai utilized digital video recorders~\cite{DBLP:conf/uss/AntonakakisABBB17}. We conclude that the introduction of \seclabel{}s might need educational campaigns that explain non-personal security risks associated with IoT devices.

%RQ3
\textit{RQ3: Differences due to demographic characteristics of the consumers.} 
The sample segmentation for the product with high perceived security risk did not show any differences in demographic factors (\Cref{tab:segmentation}). 
For the product with low perceived security risk, the segmentation showed statistically significant differences in terms of age, vocational qualification, and income.
The groups with higher age and higher income assigned a higher importance to the \seclabel{} attributes (\Cref{tab:segmentation2}).
This could indicate that this type of consumers may invest more in sustainable security of their IoT devices. 
In contrast, the group with the youngest age and lowest income seems not to care much about security updates.
We conclude that for products with a high perceived security risk, demographics may play a minor role, whereas for the products with lower perceived security risk, younger population with lower income may prefer cheaper products with lower security.

%RQ4
\textit{RQ4: Differences due to security behavior intentions, privacy concerns, and security risk perception of the consumers.} 
While the segmentation of the product category with the high perceived security risk showed only minor differences between the groups, the groups for the product category with the low perceived security risk range from very low to very high importance of the \seclabel{}.
There, the respondents with higher security behavior intentions and privacy concerns exhibit a higher preference for the attributes of the \seclabel{}. 
For both product categories, a higher security risk perception positively relates to the higher importance of the \seclabel{}'s attributes.
This indicates that consumers with a higher sensitivity for security risks and privacy concerns may assign a higher importance to the \seclabel{}s.

\subsubsection*{Limitations and Future Work}
\label{sec:discussion:limitations}

The results of the user study have the usual limitations of conjoint analysis studies.
For the comparison of products, we had to limit the number of product attributes and discretize attribute levels.
Also, some product attributes, such as product design, were discarded due to impracticality of the specification of attribute levels.
Although we specified product attributes and attribute levels based on best practices and empirical evidence, a product profile cannot fully represent all factors that may influence buying decisions, such as product presentation, packaging, advertisement campaigns, and consumer ratings.
Usability aspects of update mechanisms might also influence the consumers' choice.
Another limitation might be due to the evaluation of stated preferences (i.e., hypothetical buying decisions).
However, the assessment of real buying decisions was infeasible as there are no suitable products that guarantee security updates.
Finally, as we ran the user study with German  respondents, the results might not be valid for other markets. 
Thus, future work is required to investigate the impact of \seclabel{}s in other countries, and with different sets of participants, products, and attributes.
As we focused on the consumers' choice in this work, future work should also consider the positions of manufacturers and policy makers.

\subsubsection*{Economic Implications}
\label{sec:discussion:econ-implications}

Security update guarantees may create additional costs for the manufacturers that will be potentially passed on to the consumers~\cite{chattopadhyay-2019-single-item}. 
Are consumers willing to bear these costs?
CBC analyses are not appropriate to precisely estimate consumers’ willingness to pay for a certain attribute. 
However, attribute levels with a high importance for consumers’ choices also indicate a willingness to pay a higher price for products with this attribute level \cite{miller-2011}. 
In addition, initial support for the willingness to pay a price premium is provided by studies that show that consumers perceive price increases as fair if they are caused by higher costs for a manufacturer \cite{kahneman-1986,bolton-2003,koschate-2016}. 
Prior works \cite{DBLP:conf/chi/NaeiniDAC19,DBLP:conf/sigecom/RedmilesMD18} also concluded that consumers accept additional costs for security depending on the product's perceived security risk.

\subsubsection*{Policy Implications}
\label{sec:discussion:implications}

Our results show that mandatory \seclabel{}s could indeed have a high influence on the consumers' choices.
The labels communicate attributes that enable non-experts to compare security properties of different IoT products intuitively during the purchase process, and thereby influence buying decisions.
The introduction of \seclabel{}s might increase security of IoT consumer products through establishing economic incentives for manufacturers to guarantee a long and timely availability of security updates, or, from another perspective, creating competitive disadvantages for the non-availability of security updates.
These labels could strengthen the state of IoT security in the long term, as unpatched IoT consumer products are a major reason behind today's IoT security incidents.

%Redmiles: User study: more than half of participant make rational decisions (increase utility), consumers behave more rational if there is a higher risk.  Confirms our results
%Ferreira: implementation of security mechanisms decreases manufacturer’s profit due to increased production costs
%Emami: (z.B. dass Konsumenten bereit sind ein Price Premium für Sicherheit zu zahlen, dass Sicherheit wichtiger ist als viele andere Produktattribute und dass dieses Verhalten auch von dem Perceived Risk der Produktkategorie abhängt

%%%%%%%%%%%%%%%%%%%%%%%%%%%%
% CONCLUSION
%%%%%%%%%%%%%%%%%%%%%%%%%%%%

\section{Conclusion}
\label{sec:conclusion}

\seclabelb{}s benefit consumers, as they decrease the probability of becoming a victim to disclosed but unpatched vulnerabilities. 
They have the potential to motivate manufacturers to invest more resources in the provision of security updates, which might lead to positive security-related changes in their business strategies.
Finally, national security will also profit from these labels, as they strengthen the security of the private IoT infrastructure, and therefore, reduce the attack surface for malicious domestic and foreign actors. 

%%%%%%%%%%%%%%%%%%%%%%%%%%%%
% REFERENCES
%%%%%%%%%%%%%%%%%%%%%%%%%%%%

% Acknowledgements:
% - Statistik-Expertin Andrea 
% - Statistik-Expertin Freya Gassmann
% - Sawtooth Software
% - Shepherd Nick Feamster 

%\vspace{2mm}
\section*{Acknowledgments}

We thank Freya Gassmann and Andrea Schankin for consulting regarding the statistical evaluation of the user study.
We further thank the anonymous reviewers for their thorough and valuable comments, as well as the shepherd of this paper, Nick Feamster, for his guidance.
This work was supported by a software grant from Sawtooth Software.

%=======
%We thank Freya Gassmann and Andrea Schankin for consulting regarding the statistical evaluation of the user study.
%We further thank the reviewers for their thorough and valuable comments, as well as the shepherd of this paper, Nick Feamster, for his guidance.
%This work was supported by a Sawtooth Software grant.
%>>>>>>> .r9011

%\newpage 
\bibliographystyle{IEEEtran}
\bibliography{bib/bib-zll-philipp,bib/bib-z3,bib/bib-iot,bib/bib-kernel,bib/bib-econ,bib/bib-econ-label,bib/bib-metrics,bib/bib-zb-seclila}

%%%%%%%%%%%%%%%%%%%%%%%%%%%%
% APPENDIX
%%%%%%%%%%%%%%%%%%%%%%%%%%%%

\appendix
%\section{Study Design and Replication}
%\label{app:replication} 
%
%\philipp{Wollen wir gesammelte Daten online stellen? Wenn ja, welche?}
%We provide all necessary details to replicate this study.
%The full survey designs and collected raw data will be published on the research group's GitHub account.  
%For the double-blind reviews of this submission, we provide a zip-file of all data for download: \url{www.link.com}
%
%In the published raw data, we withhold sensitive information (e.g., respondents' IP addresses) to conform to existing privacy legislation.

\section{Further Material}
%\label{app:security-risk-scale} 

\subsection{Marketing Research Scales of Prestudy 1}
\label{app:prestudy1:scales} 

The scales below were used in Prestudy 1 to compare the perception of product categories.
%All items were measured on a 7-point Likert scale.

%\zina{Prestudy 1: we should present these scales in Appendix, with references and also explain if we adapted any of them (removed items, rephrased)}
\begin{packed_item}

\item Attitude towards a product category adapted from Martin et al.~\cite{martin-2005} 
% Original: 5pt, example: sport shoes, attitude towards parent brand
%(5pt, 1 = not at all favorable/likable/pleasing to 5 = very favorable/likable/pleasing)
\begin{itemize}
\item How favorable are [product]s? \\ \textit{(1 = `not at all favorable' to 7 = `very favorable')}
\item How likable are [product]s? \\ \textit{(1 = `not at all likable' to 7 = `very likable')}
\item How pleasing are [product]s? \\ \textit{(1 = `not at all pleasing' to 7 = `very pleasing')}
\end{itemize}

\item Involvement with a product category, Mathwick and Rigdon~\cite{mathwick-2004}. \\
%Scale: 7-point semantic differential.
%\zina{the general statement must be presented, I guess it's something like "Please indicate your attitude to product X ..."}
`[Product]s...'
%Original: 7pt, Product involvement
\begin{itemize}
\item 1 = `Mean nothing to me' to 7 = `mean a lot to me' 
\end{itemize}
`I find [product]s...'
\begin{itemize}
\item 1 = `worthless' to 7 = `valuable'
\item 1 = `boring' to 7 = `interesting'
\item 1 = `exciting' to 7 = `unexciting' 
\item 1 = `fascinating' to 7 = `mundane'
\item 1 = `involving' to 7 = `uninvolving'
\end{itemize}

\item Consumption motives, Wakefield and Inman \cite{wakefield-2003}\\
`[Product]s are used for ...'
% Original: 7pt, functional/hedonic context of the consumption occasion
\begin{itemize}
\item 1 = `Practical purposes' to 7 = `Just for Fun'
\item 1 = `Purely functional' to 7 = `Pure enjoyment'
\item 1 = `For a routine need' to 7 = `For pleasure'
\end{itemize}

\item Desirability to possess a product adapted from Lei et al.~\cite{lei-2012}
% Original: Brand desirability
\begin{itemize}
\item ~[Product]s are… \textit{(1 = `not at all desirable' to 7 = `very desirable')}
\end{itemize}

\item Purchase intention adapted from MacKenzie et al.\  \cite{mackenzie-1986}
\textit{(1 = `very low' to 7 = `very high')}
% Original: ???
\begin{itemize}
\item If I were going to buy a [product], the probability of buying this model is...
\item The probability that I would consider buying this [product] is...
\item The likelihood that I would purchase this [product] is...
\end{itemize}

\end{packed_item}

\subsection{Quality Analysis of Perceived Security Risk Scale}
\label{app:details:pca} 

We assessed the quality criteria of the perceived security risk scale based on the results of the conjoint questionnaire, where the group of 731 respondents (denoted as PC1) evaluated a product category with a high perceived security risk, and the group of 735 respondents (denoted as PC2) evaluated a product category with a low perceived security risk.

First, we evaluated the convergent validity of the perceived security risk scale using a principal component analysis (PCA) with Varimax rotation, in order to prove the dimensionality. % as suggested by Homburg and Giering \cite{homburg1998}.
For the PCA, the Bartlett test indicated significant correlations, the Kaiser-Meyer-Olkin (KMO) measure verified the sampling adequacy with KMO = .898 for PC1 and KMO = .917 for PC2, and all measures of sampling adequacy (MSA) values for individual items were greater or equal than .803 for PC1 and .840 for PC2. 
Thus, KMO and MSA exceed the acceptable limit of .500 as proposed by Field~\cite{field2013-spss}.
%\zina{I don't understand what this text means:}
%Field \cite{field2013-spss} suggests an alternative eigenvalue cutoff (e.g., Jolliffe’s eigenvalue \cite{jolliffe1972}) instead of the Kaiser’s criterion, if the sample size exceeds 250 and the average communality is greater than .6, which applies to this case. 
Field also suggests an eigenvalue cutoff, if the sample size exceeds 250 and the average communality is greater than .6.
As both conditions apply to our case, we used an eigenvalue cutoff of .7, as proposed by Jolliffe \cite{jolliffe1972}. 
%This procedure is very similar to Rust et al.\ \cite{rust2004}, and helps to provide the best trade-off between parsimony and interpretability concerning the scale. 
This procedure is similar to Rust et al.\ \cite{rust2004}, and helps to provide the best trade-off between parsimony and interpretability concerning the scale. 

The factors of PC1 and PC2 are shown in \Cref{tab:scale-items}.
For PC1, the PCA with Varimax rotation for the 13 perceived security items revealed four factors for perceived security (variance explained = 77.94\%). 
For PC2, the PCA with Varimax rotation revealed three factors (variance explained = 77.14\%). 
Factor loadings of PC1 and PC2 were greater or equal to .464 (should be \textgreater .400) \cite{hair2010}.  
For PC1, all items loaded the highest on the factor (i.e., `general', `privacy', `physical', or `financial risk') on which they were supposed to load, making all factors easy to interpret. 
Thus, the four factors aligned closely with the four perceived security dimensions that we derived from previous research.
For PC2, the first seven items formed a composite factor (`general/privacy risk'), while items 8 to 10 loads on `physical risk', and items 11 to 13 loaded on `financial risk'. The reasons for the  loading of the general and privacy risks to a single factor are not clear, such that the scale may require further refinement in the future.

%In the statistical analysis, the perceived security risk scale showed excellent values ($\geq.982$) in terms of the tau-equivalent reliability (also known as Cronbach's alpha), a measure that defines the inner consistency of a scale.

Second, we evaluated the perceived security risk scale in terms of the Cronbach's alpha, a measure that defines the inner consistency of a scale, and the item-to-total correlation for all dimensions. 
%We removed item \#9 from the dimension `physical' because it showed an inter-item correlation of less than the critical threshold of .500 \cite{field2013-spss,homburg1998}. 
For all factors of PC1 and PC2, Cronbach's alpha was above the recommended threshold of .700 (\textgreater .804) and item-to-total correlation higher than the threshold of .500 (\textgreater  .524), indicating high reliability and internal consistency of the perceived security risk scale \cite{field2013-spss}.

% Table: Results Prestudy 2
%% Product attributes: Weather Station
\begin{table}
  \centering
  \caption{Considered product attributes ($n=29$) and their dual-questioning scores (higher score values indicate a higher perceived importance for this attribute)}
  \label{tab:app:attributes-scores}
  \renewcommand{\arraystretch}{0.8}
  \begin{tabular}[htb]{@{}c@{\hspace{2ex}} @{}L{6cm}@{\hspace{2ex}} @{}r@{\hspace{2ex}} r@{}}
    \toprule
    \multicolumn{4}{c}{\bf Smart Home Camera} \\
    \midrule
    \textit{Rank} & \textit{Product Attribute}  & $\mu$ & $\sigma$  \\
    \midrule
    1. & Price & 31.76 & 12.83  \\
    2. & Solar panel for energy generation  & 26.66 & 12.01  \\
    3. & Battery lifetime & 26.31 & 11.33 \\
    4. & Precision level of measurements & 23.72 & 10.98  \\
    5. & Rain and wind measurements & 23.69 & 10.99  \\
    6. & Expandability for multiple rooms & 23.48 & 13.06 \\
    7. & Max. wireless range  & 23.34 & 10.17  \\
    8. & Warning feature (e.g., storm) & 22.66 & 10.59  \\
    9. & How many days of local weather forecast & 22.14 & 8.80 \\
    10. & Measurement rate & 19.14 & 10.15 \\
    11. & Brand/ manufacturer & 18.28 & 10.12 \\
    12. & Alarm upon preset threshold exceedance  & 18.03 & 10.69 \\
    13. & Seal of technical approval & 17.76 & 10.14 \\
    14. & Environmental label & 17.14 & 12.23 \\
    15. & Material of casing & 16.62 & 10.45 \\
    16. & Color of product & 16.03 & 11.69 \\
    17. & Alexa compatible & 12.34 & 9.93 \\
    18. & Accessibility label (fictive) & 11.17 & 6.70 \\
    \midrule
    \multicolumn{4}{c}{\bf Smart Weather Station} \\
    \midrule
    \textit{Rank} & \textit{Product Attribute}  & $\mu$ & $\sigma$  \\
    \midrule
    1. & Price & 30.59 & 12.69  \\
    2. & Video resolution & 26.79 & 12.22   \\
    3. & Field of view & 26.17 & 11.60 \\
    4. & Video frame rate & 24.41 & 11.21   \\
    5. & Zoom function & 23.10 & 11.85  \\
    6. & Energy consumption & 22.59 & 10.53 \\
    7. & Face recognition & 22.10 & 12.19  \\
    8. & Night vision mode & 21.79 & 9.60 \\
    9. & Timing function for recordings & 20.86 & 9.16 \\
    10. & Type of power supply & 19.24 & 8.59 \\
    11. & Brand/ manufacturer & 18.48 & 11.92 \\
    12. & Seal of technical approval & 18.48 & 12.21 \\
    13. & Material of encasement & 18.41 & 10.62 \\
    14. & Environmental label & 17.38 & 11.68 \\
    15. & SD card slot & 17.21 & 9.69 \\
    16. & Color of product & 14.52 & 10.21 \\
    17. & Accessibility label (fictive) & 14.14 & 8.73 \\
    18. & Alexa compatible & 11.21 & 9.19 \\
    \bottomrule
    & \\
  \end{tabular}
\end{table}

\subsection{Consistent Akaike Information Criterion}
\label{app:details:akaike} 

The consistent Akaike information criterion (CAIC) is recommended by Sawtooth Software as measure to decide on the number of reasonable segments  \cite{sawtooth-latent-class} for latent-class analysis.
Due to the sample size (cf.\ \Cref{sec:study:conjoint:data}), we were interested in splitting the respondents into either 2 or 3 groups, for which the latent-class analysis provided the corresponding proposals.
Each segmentation proposal has a CAIC score.
Smaller values of CAIC are desired \cite{sawtooth-latent-class}. % Sawtooth Latent Class Paper
For the segmentation of the respondents that evaluated the smart home camera, the CAIC score was  14,162 for 2 groups and 14,307 for 3 groups. 
For the segmentation of the respondents that evaluated the smart weather station, the CAIC score was 14,471 for 2 groups and 14,260 for 3 groups.
In summary, for the smart home cameras, the smaller CAIC score indicated 3 groups, while for smart weather, the smaller CAIC score indicated 2 groups.

%P1: 
% 1: 13909.91477
% 2: 14161.73995 <--
% 3: 14306.84777
%
%P2: 
% 1: 15247.75139
% 2: 14471.20685
% 3: 14260.41336 <--

% Figure: Label Examples
    \begin{figure}[t]
        \centering
	{%
	\setlength{\fboxsep}{5pt}%
	\setlength{\fboxrule}{0.5px}%
	\fbox{\includegraphics[trim=0 0 0 0, clip, width=0.34\textwidth]{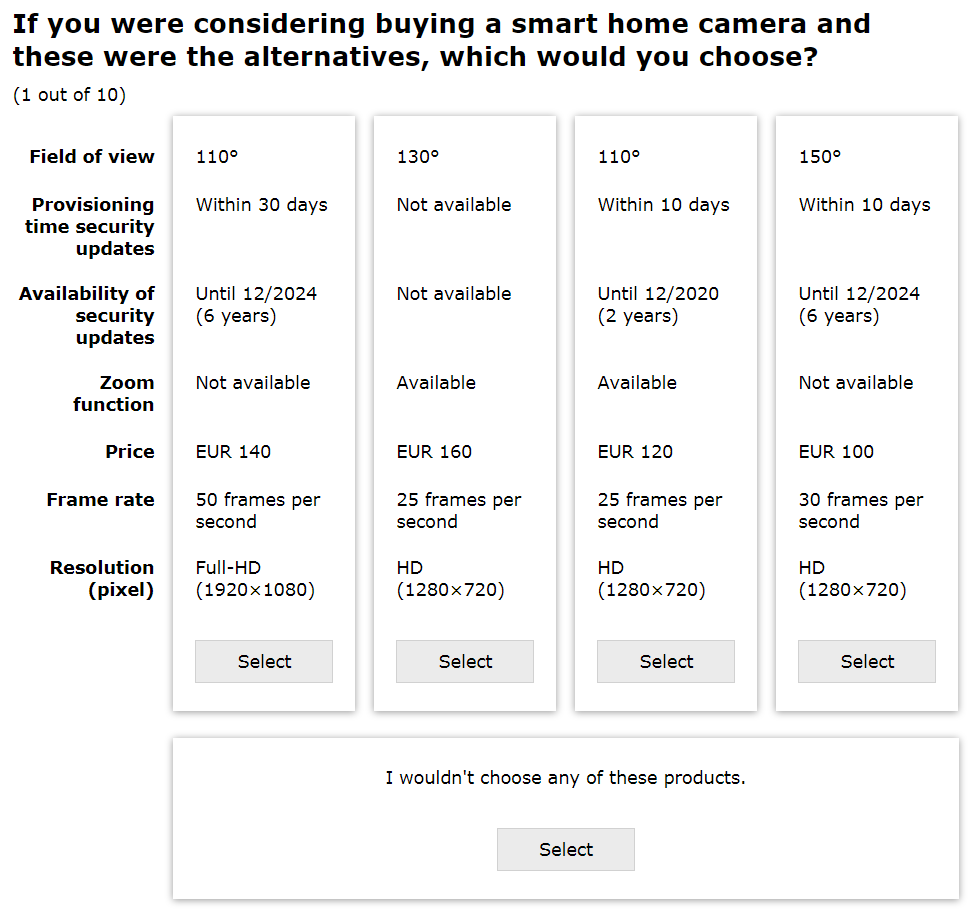}}%
	}%
        \caption{Screenshot of a choice task in the conjoint questionnaire (translated to English).}
	\label{fig:choice-task}
    \end{figure}

% Table: Prestudy 1 (Scores)
\begin{table}
  \centering
  \caption{Mean $\mu$ and standard deviation $\sigma$ from Prestudy~1 ($n=30$)  sorted by perceived security risk.} % measured on 7-pt Likert scales and 
  \label{tab:app:results-experiment1}%
  \renewcommand{\arraystretch}{0.6}
%  \begin{tabular}[htb]{@{}l @{}r@{\hspace{2ex}} r@{\hspace{2ex}}r@{\hspace{2ex}}r@{\hspace{2ex}}r@{\hspace{2ex}} r@{\hspace{2ex}}r@{\hspace{2ex}}r }
  \begin{tabular}[htb]{@{}l @{}c@{\hspace{2ex}}r@{\hspace{2ex}}r@{\hspace{2ex}}r@{\hspace{2ex}} r@{\hspace{2ex}}r@{\hspace{2ex}}r@{}}
    \toprule
    \rule{0pt}{14ex}% 
%    Product Category  & \tabrotate{Ownership [\%]} & \tabrotate{Estimated Price [Euro]} & \tabrotate{Attitude} & \tabrotate{Consumption Motive} & \tabrotate{Purchase Intention} & \tabrotate{Involvement} & \tabrotate{Desirability} & \tabrotate{Perceived Security Risk} \\
   \bf Product Category   & & \tabrotate{\pbox{2.2cm}{\bf Perceived {Security} Risk}} & \tabrotate{\bf Attitude} & \tabrotate{\pbox{1.5cm}{\bf Consumption Motive}} & \tabrotate{\pbox{1.4cm}{\bf Purchase Intention}} & \tabrotate{\bf Involvement} & \tabrotate{\bf Desirability}  \\
    \cmidrule(r){1-2}
    \cmidrule(r){3-3}
    \cmidrule(){4-8}

\multirow{2}{*}{Smart Alarm System}
				 				& $\mu$		&	5.76 &	4.77 &	1.74 &	3.37 &	4.26 &	3.90 \\
								& $\sigma$	&	1.33 &	1.17 &	1.01 &	1.57 &	1.13 &	1.47 \\
    \cmidrule(r){1-2}
    \cmidrule(r){3-3}
    \cmidrule(){4-8}
\multirow{2}{*}{Smart Door Lock} 					
								& $\mu$		&	5.65 &	3.16 &	2.42 &	2.39 &	3.22 &	2.70 \\
								& $\sigma$	&	1.63 &	1.38 &	1.43 &	1.55 &	1.28 &	1.34 \\
    \cmidrule(r){1-2}
    \cmidrule(r){3-3}
    \cmidrule(){4-8}
\multirow{2}{*}{Smart Home Camera}				
								& $\mu$		&	5.49 & 	3.86 &	3.44 &	3.02 &	3.56 &	3.57 \\
								& $\sigma$	&	1.41 &	1.82 &	1.73 &	1.60 &	1.41 &	1.59 \\
    \cmidrule(r){1-2}
    \cmidrule(r){3-3}
    \cmidrule(){4-8}
\multirow{2}{*}{Smart Smoke Detector~~	}		
								& $\mu$		&	4.12 & 	5.16 &	1.74 &	4.27 &	4.52 &	4.27 \\
								& $\sigma$	&	1.55 &	1.34 &	1.03 &	1.79 &	1.30 &	1.55 \\
    \cmidrule(r){1-2}
    \cmidrule(r){3-3}
    \cmidrule(){4-8}
\multirow{2}{*}{Smart Thermostat} 				
								& $\mu$		&	3.67 & 	5.24 &	1.72 &	4.52 &	4.52 &	4.67 \\
								& $\sigma$	&	1.61 &	1.38 &	0.97 &	1.69 &	1.20 &	1.27 \\
    \cmidrule(r){1-2}
    \cmidrule(r){3-3}
    \cmidrule(){4-8}
\multirow{2}{*}{Smart Light Bulb} 					
								& $\mu$		&	3.40 &	3.90 &	4.39 &	3.24 &	3.31 &	3.33 \\
								& $\sigma$	&	1.45 &	1.48 &	1.63 &	1.61 &	1.41 &	1.56 \\
    \cmidrule(r){1-2}
    \cmidrule(r){3-3}
    \cmidrule(){4-8}
\multirow{2}{*}{Smart Vacuum Robot }				
								& $\mu$		&	3.10 &	4.76 &	2.30 &	4.08 &	4.12 &	3.97 \\
								& $\sigma$	&	1.67 &	1.84 &	1.38 &	2.09 &	1.54 &	1.87 \\
    \cmidrule(r){1-2}
    \cmidrule(r){3-3}
    \cmidrule(){4-8}
\multirow{2}{*}{Smart Weather Station} 			
								& $\mu$		&	2.98 & 	4.38 &	2.90 &	3.51 &	3.75 &	3.67 \\
								& $\sigma$	&	1.51 &	1.83 &	1.47 &	1.87 &	1.63&	1.69 \\

    \bottomrule
  \end{tabular}
\end{table}

%Smart Weather Station & 	23 &	 104.57 &	4.38 &	2.90 &	3.51 &	3.75 &	3.67 &	2.98\\
%Smart Vacuum Robot &	27 &	282.50 &	4.76 &	2.30 &	4.08 &	4.12 &	3.97 &	3.10\\
%Smart Light Bulb &	13 &	68.23 &	3.90 &	4.39 &	3.24 &	3.31 &	3.33 &	3.40 \\
%Smart Thermostat & 	33 &	87.37 &	5.24 &	1.72 &	4.52 &	4.52 &	4.67 &	3.67\\
%Smart Smoke Detector~~ & 10 &	78.77 &	5.16 &	1.74 &	4.27 &	4.52 &	4.27 &	4.12\\
%Smart Home Camera	& 17 &	168.00 &	3.86 &	3.44 &	3.02 &	3.56 &	3.57 &	5.49\\
%Smart Door Lock & 7	& 190.10 &	3.16 &	2.42 &	2.39 &	3.22 &	2.70 &	5.65 \\
%Smart Alarm System &	10 &	520.80 &	4.77 &	1.74 &	3.37 &	4.26 &	3.90 &	5.76\\

% Table: Posthoc Tests
%% Product attributes: Weather Station
\begin{table}
  \centering
  \caption{Post-hoc tests (Tukey HSD) for segmentation.}
  \label{tab:app:posthocs}
    \renewcommand{\arraystretch}{0.6}
    \setlength{\tabcolsep}{4pt}
  \begin{tabular}[htb]{L{2.3cm} c r r c r}
    \toprule
    \multicolumn{6}{c}{\bf Smart Weather Station} \\
    \midrule
    \textit{} & \textit{Groups}  &   \textit{MD\textsuperscript{1}} & $p$ & 95\%-CI\textsuperscript{2} & d \\
    \midrule
    \multirow{3}{*}{Age}
    							& 1 vs 2 & $2.14$ & $0.14$ & $[-0.51, 4.80]$ & $0.19$  \\
    							& 1 vs 3 & $4.12$ & $0.00$ & $[1.68, 6.57]$ &  $0.37$  \\
    							& 2 vs 3 & $1.98$ & $0.11$ & $[-0.32, 4.28]$ &  $0.17$  \\
    \midrule
    \multirow{3}{*}{Monthly net income}
    							& 1 vs 2 & $146.95$ & $0.50$ & $[-158.63, 452.53]$ &  $0.11$  \\
    							& 1 vs 3 & $322.04$ & $0.02$ & $[37.85, 606.49]$ &  $0.26$  \\
    							& 2 vs 3 & $175.09$ & $0.26$ & $[-85.76, 435.94]$ &  $0.15$  \\
    \midrule
    \multirow{3}{*}{Privacy Concerns}
    							& 1 vs 2 & $-0.03$ & $0.97$ & $[-0.37, 0.31]$ &  $0.02$  \\
    							& 1 vs 3 & $0.28$ & $0.09$ & $[-0.03, 0.59]$ &  $0.20$  \\
    							& 2 vs 3 & $0.31$ & $0.03$ & $[0.02, 0.60]$ &  $0.22$  \\
    \midrule
    \multirow{3}{2.3cm}{Security behavior intentions}
    							& 1 vs 2 & $0.03$ & $0.89$ & $[-0.12, 0.18]$ &  $0.06$  \\
    							& 1 vs 3 & $0.23$ & $0.00$ & $[0.09, 0.36]$ &  $0.39$  \\
    							& 2 vs 3 & $0.20$ & $0.00$ & $[0.07, 0.33]$ &  $0.47$  \\
    \midrule
    \multirow{3}{2.3cm}{Perceived security risk}
    							& 1 vs 2 & $-0.08$ & $0.84$ & $[-0.41, 0.25]$ &  $0.05$  \\
    							& 1 vs 3 & $0.56$ &  $0.00$ & $[0.25, 0.86]$ &  $0.36$  \\
    							& 2 vs 3 & $0.63$ & $0.00$ & $[0.35, 0.92]$ &  $0.31$  \\
%    \midrule
%    							& 1 vs 2 & $2.14, p = 0.14 , 95\%-CI[-0.51, 4.80], d = 0.19$  \\
%    Age						& 1 vs 3 & $4.12, p = 0.00, 95\%-CI[1.68, 6.57], d = 0.37$  \\
%    							& 2 vs 3 & $1.98, p = 0.11, 95\%-CI[-0.32, 4.28], d = 0.17$  \\
%    \midrule
%    							& 1 vs 2 & $146.95, p = 0.50, 95\%-CI[-158.63, 452.53], d = 0.11$  \\
%    Monthly net income			& 1 vs 3 & $322.04, p = 0.02, 95\%-CI[37.85, 606.49], d = 0.26$  \\
%    							& 2 vs 3 & $175.09, p = 0.26, 95\%-CI[-85.76, 435.94], d = 0.15$  \\
%    \midrule
%    							& 1 vs 2 & $-0.03, p = 0.97, 95\%-CI[-0.37, 0.31], d = 0.02$  \\
%    Privacy Concerns			& 1 vs 3 & $0.28, p = 0.09, 95\%-CI[-0.03, 0.59], d = 0.20$  \\
%    							& 2 vs 3 & $0.31, p = 0.03, 95\%-CI[0.02, 0.60], d = 0.22$  \\
%    \midrule
%    							& 1 vs 2 & $0.03, p = 0.89, 95\%-CI[-0.12, 0.18], d = 0.06$  \\
%    Security behavior intentions	& 1 vs 3 & $0.23, p = 0.00, 95\%-CI[0.09, 0.36], d = 0.39$  \\
%    							& 2 vs 3 & $0.20, p = 0.00, 95\%-CI[0.07, 0.33], d = 0.47$  \\
%    \midrule
%    							& 1 vs 2 & $-0.08, p = 0.84, 95\%-CI[-0.41, 0.25], d = 0.05$  \\
%    Perceived security risk		& 1 vs 3 & $0.56, p = 0.00, 95\%-CI[0.25, 0.86], d = 0.36$  \\
%    							& 2 vs 3 & $0.63, p = 0.00, 95\%-CI[0.35, 0.92], d = 0.31$  \\
    \bottomrule
  \end{tabular}
     \begin{tablenotes}
      \footnotesize
      \item Notation: \textsuperscript{1}mean difference, \textsuperscript{2}95\%-confidence interval
  \end{tablenotes}
\end{table}

% Table: Security Risk Scale
\begin{table*}
  \centering
  \captionof{table}{Perceived security risk scale (translated to English) with quality criteria.}
  \label{tab:scale-items}
  \begin{tabular}{@{}r L{2cm}@{\hspace{2ex}}L{6.5cm} R{0.9cm}@{\hspace{2ex}}R{0.9cm}@{\hspace{2ex}}R{1.4cm}@{\hspace{4ex}}R{0.9cm}@{\hspace{2ex}}R{0.9cm}@{\hspace{2ex}}R{1.4cm}@{}}
    \toprule
    \multirow{2}{*}{\bf \#} & \bf Risk Category from \cite{DBLP:journals/ijmms/FeathermanP03} & \multirow{2}{*}{\textbf{Item}} & \multicolumn{3}{C{3.7cm}}{\textbf{Product Category PC1} \pbox{3.7cm}{(4 factors, $n=731$)}}  &\multicolumn{3}{C{3.6cm}}{\textbf{Product Category PC2} \pbox{3.6cm}{(3 factors, $n=735$)}}\\
    \midrule
     &  & If a third party takes unauthorized control over [product], there is a high risk that... & \it Factor loading  &  \it Item to Total & \it {Cronbach's} alpha & \it  Factor loading  & \it  Item to Total & \it {Cronbach's} alpha\\
    \midrule
    
    1 & \multirow{4}{*}{General} & ...the consequences are severe.								& .769 & .715 & \multirow{4}{*}{.891} 	& .784 & .780 & \multirow{8}{*}{.938} \\ 
    2 & & ...it leads to high potential of abuse. 													& .835 & .790 &   						& .817 & .812 &  \\ 
    3 & & ...it is used for criminal purposes. 														& .750 & .768 &   						& .792 & .795 & \\ 
    4 & & ...a serious security threat exists.														& .776 & .783 &   						& .822 & .816 & \\ 

    \cmidrule(r){1-6}
    5 & \multirow{3}{*}{Privacy} & ...this has a serious impact on privacy. 							& .583 & .596 &  							& .787 & .809 &  \\ 
    6 & & ...they access personal information. 													& .840 & .774 & .826  					& .742 & .800 & \\ 
    7 & & ...it steals private data.																& .826 & .710 &  							& .702 & .761 &  \\ 

    \midrule
    8 & \multirow{4}{*}{Physical} & ...the health of its owners or other people is at risk.	 			& .885 & .716 & \multirow{4}{*}{.804}	& .881 & .803 & \multirow{4}{*}{.876}\\ 
    9 & & ...the safety of its owners or other people is at risk. 										& .464 & .524 &   					& .608 & .688 & \\ 
    \multirow{2}{*}{10} & & ...it has harmful consequences to the physical integrity of its owner or other people.			& \multirow{2}{*}{.875} & \multirow{2}{*}{.733} &  							& \multirow{2}{*}{.882} & \multirow{2}{*}{.802} & \\ 

    \midrule
    11 & \multirow{3}{*}{Financial} & ...the owner suffers financial losses. 							& .808 & .756 &   						& .848 & .797 & \\ 
    12 & & ...it is misused for crimes involving financial loss.										& .799 & .737 & .886  					& .716 & .781 & .906\\ 
    13 & & ...it leads to financial loss. 															& .854 & .841 & 							& .846 & .863 &  \\ 

%    \midrule
%    \textit{14} & \multirow{4}{*}{\textit{Performance}} & \textit{...the [product] does not work properly anymore.}* &   & & \\
%    %\multicolumn{3}{L{4cm}}{\multirow{4}{5cm}{\textit{We surveyed these items in  Prestudy 1 but then excluded them due to parsimony as we realized that there is a high risk on performance for all product categories, which does not dependent on security-criticalness.}}}  \\ 
%    \textit{15} & & \textit{...the [product] can no longer be controlled by the user.}*  &  &  &  \\ 
%    \textit{16} & & \textit{...the [product]’s features can no longer be used as intended.}*  &  &  & \\ 
%    \textit{17} & & \textit{...the functionality of [product] is limited.}*  &  &  &  \\ 
    \bottomrule
  \end{tabular}
  \begin{tablenotes}
      \footnotesize
      \item Items measured on 7-point Likert scale ranging from 1 = `strongly disagree' to 7 = `strongly agree'.
  \end{tablenotes}
\end{table*}

% Table: Prestudy 1 (Significances)
\begin{table*}
  \centering
  \caption{Statistical significance and effect sizes (Cohen's $d_z$) of central product attitudes, purchase intention, and security risk perception for pairs of product categories (Prestudy 1).}
  \label{tab:app:significance}%
   \setlength{\tabcolsep}{8pt}
  \begin{tabular}[htb]{@{}L{1.9cm}@{\hspace{1ex}}L{2.1cm}@{\hspace{2ex}}r@{\hspace{0ex}}l@{\hspace{1ex}}r@{\hspace{3ex}}r@{\hspace{0ex}}l@{\hspace{1ex}} r@{\hspace{3ex}}r@{\hspace{0ex}}l@{\hspace{1ex}}r@{\hspace{3ex}}r@{\hspace{0ex}}l@{\hspace{1ex}}r@{\hspace{3ex}}r@{\hspace{0ex}}l@{\hspace{1ex}}r@{\hspace{3ex}}r@{\hspace{0ex}}l@{\hspace{1ex}}r@{}}
    \toprule
%    \rule{0pt}{4ex}% 
      \multirow{3.5}{*}{\pbox{1.4cm}{\bf Product Category 1}}  
      & \multirow{3.5}{*}{\pbox{1.4cm}{\bf Product Category 2}} 
      & \multicolumn{3}{c}{\multirow{2}{*}{\bf Attitude}} 
      & \multicolumn{3}{C{1.5cm}}{\bf Consumption Motive} 
      &  \multicolumn{3}{C{1.5cm}}{\bf Purchase Intention} 
      & \multicolumn{3}{c}{\multirow{2}{*}{\bf Involvement}} 
      & \multicolumn{3}{c}{\multirow{2}{*}{\bf Desirability}} 
      &  \multicolumn{3}{C{1.7cm}}{\bf Perceived {Security} Risk} \\
   % \cmidrule(r){1-2}
    \cmidrule(r){3-5}
    \cmidrule(r){6-8}
    \cmidrule(r){9-11}
    \cmidrule(r){12-14}
    \cmidrule(r){15-17}
    \cmidrule(){18-20}
    & & \multicolumn{2}{r}{$t(29)$} & $d_z$  & \multicolumn{2}{r}{$t(29)$} & $d_z$  & \multicolumn{2}{r}{$t(29)$} & $d_z$  & \multicolumn{2}{r}{$t(29)$} & $d_z$  & \multicolumn{2}{r}{$t(29)$} & $d_z$  & \multicolumn{2}{r}{$t(29)$} & $d_z$ \\
    \cmidrule(r){1-2}
    \cmidrule(r){3-5}
    \cmidrule(r){6-8}
    \cmidrule(r){9-11}
    \cmidrule(r){12-14}
    \cmidrule(r){15-17}
    \cmidrule(){18-20}

Door lock & Light bulb  					&	-2.266&\textsuperscript{*}  	&	-0.414	&	-4.855&\textsuperscript{***} 	& 	-0.886 	&	-2.762&\textsuperscript{*}	&	-0.504 	&	-0.338& 						&	-0.062 	&	-2.129&\textsuperscript{*}	&	-0.389 	& 	~~~~5.894&\textsuperscript{***} 	&	1.076 \\
Door lock & Home Camera 				&	-2.371&\textsuperscript{*}	&	-0.433 	&	-4.237&\textsuperscript{***} 	&	-0.773 	&	-2.828&\textsuperscript{**} 	&	-0.516 	&	-1.970& 						&	-0.360	&	-3.496&\textsuperscript{**} 	&	-0.638 	&	0.637& 						&	0.116 \\
Door lock & Smoke detector 				&	-7.062&\textsuperscript{***} 	&	-1.289 	&	2.088&\textsuperscript{*}		&	0.381 	&	-5.130&\textsuperscript{***} 	&	-0.937 	&	-4.224&\textsuperscript{***} 	&	-0.771 	&	-4.419&\textsuperscript{***} 	&	-0.807 	&	5.288&\textsuperscript{***} 	&	0.965 \\
Door lock & Thermostat 					&	-6.519&\textsuperscript{***} 	& 	-1.190	&	2.327&\textsuperscript{*}		&	0.425 	&	-5.884&\textsuperscript{***} 	&	-1.074 	&	-4.811&\textsuperscript{***} 	&	-0.878 	&	-5.662&\textsuperscript{***} 	&	-1.034	&	5.788&\textsuperscript{***} 	&	1.057 \\
Door lock & Vacuum robot 				&	-4.636&\textsuperscript{***} 	&	-0.846 	&	.388& 						&	0.071	&	-4.809&\textsuperscript{***} 	&	-0.878 	&	-2.783&\textsuperscript{**} 	&	-0.508 	&	-3.532&\textsuperscript{**} 	&	-0.645 	&	6.462&\textsuperscript{***} 	&	1.180 \\
Door lock & Weather station 				&	-3.600&\textsuperscript{**} 	&	-0.657 	&	-1.563& 						&	-0.285 	&	-3.063&\textsuperscript{**}	&	-0.559 	&	-1.891& 						&	-0.345 	&	-3.057&\textsuperscript{**} 	&	-0.558 	&	7.298&\textsuperscript{***} 	&	1.332 \\ 
Door lock & Alarm system 					&	-8.148&\textsuperscript{***} 	&	-1.488 	&	2.411&\textsuperscript{*}		&	0.440 	&	-4.463&\textsuperscript{***} 	&	-0.815 	&	-5.442&\textsuperscript{***} 	&	-0.994 	&	-4.966&\textsuperscript{***}	&	-0.907 	&	-0.928& 						&	-0.169 \\

    \cmidrule(r){1-2}
    \cmidrule(r){3-5}
    \cmidrule(r){6-8}
    \cmidrule(r){9-11}
    \cmidrule(r){12-14}
    \cmidrule(r){15-17}
    \cmidrule(){18-20}

Light bulb & 	Home Camera 				&	0.134& 						&	0.024 	&	2.218&\textsuperscript{*}		&	0.405 	&	0.677& 						&	0.124 	& 	-0.829 &						&	-0.151 	&	-0.646&  						&	-0.118 	&	-7.024&\textsuperscript{***} 	&	-1.282 \\
Light bulb & 	Smoke detector 				&	-3.621&\textsuperscript{**}  	&	-0.661	&	6.662&\textsuperscript{***} 	&	1.216 	&	-2.425&\textsuperscript{*}	&	-0.443 	&	-3.389&\textsuperscript{**} 	&	-0.619 	&	-2.328&\textsuperscript{*}	&	-0.425 	&	-2.397&\textsuperscript{*}	&	-0.438 \\
Light bulb & 	Thermostat 					&	-4.036&\textsuperscript{***} 	&	-0.737 	&	7.515&\textsuperscript{***} 	&	1.372 	&	-3.443&\textsuperscript{**} 	&	-0.629 	&	-3.931&\textsuperscript{***} 	&	-0.718 	&	-3.641&\textsuperscript{**} 	&	-0.665 	&	-1.164& 						&	-0.213 \\
Light bulb & 	Vacuum robot 				&	-2.780&\textsuperscript{**}  	&	-0.508	&	5.917&\textsuperscript{***} 	&	1.080 	&	-2.281&\textsuperscript{*}	&	-0.416 	&	-2.710&\textsuperscript{*}	&	-0.495 	&	-1.698& 						&	-0.310	&	1.339& 						&	0.244 \\
Light bulb & 	Weather station 				&	-1.677& 						&	-0.306 	&	4.925&\textsuperscript{***} 	&	0.899 	&	-0.958& 						&	-0.175	&	-1.637& 						&	-0.299 	&	-1.095& 						&	-0.200	&	1.948& 						&	0.356 \\
Light bulb & 	Alarm system 				&	-2.991&\textsuperscript{**}  	&	-0.546	&	7.385&\textsuperscript{***} 	&	1.348 	&	-0.432& 						&	-0.079 	&	-3.372&\textsuperscript{**} 	&	-0.616 	&	-1.876& 						&	-0.343 	&	-7.225&\textsuperscript{***} 	&	-1.319 \\

    \cmidrule(r){1-2}
    \cmidrule(r){3-5}
    \cmidrule(r){6-8}
    \cmidrule(r){9-11}
    \cmidrule(r){12-14}
    \cmidrule(r){15-17}
    \cmidrule(){18-20}

Home Camera & 	Smoke detector 			&	-3.876&\textsuperscript{**}  	&	-0.708 	&	4.765&\textsuperscript{***} 	&	0.870 	&	-3.850&\textsuperscript{**} 	&	-0.703 	&	-3.148&\textsuperscript{**}  	&	-0.575 	&	-2.104&\textsuperscript{*}	&	-0.384	&	4.348&\textsuperscript{***}	&	0.794 \\
Home Camera & 	Thermostat 				&	-4.200&\textsuperscript{***} 	&	-0.767 	&	5.208&\textsuperscript{***} 	&	0.951 	&	 -5.324&\textsuperscript{***} 	&	-0.972	&	-3.707&\textsuperscript{**} 	&	-0.677 	&	-3.171&\textsuperscript{**} 	&	-0.579 	&	5.799&\textsuperscript{***} 	&	1.059 \\
Home Camera & 	Vacuum robot			&	-2.631&\textsuperscript{*}	&	-0.480 	&	3.301&\textsuperscript{**} 	&	0.603 	&	-3.353&\textsuperscript{**} 	&	-0.612 	&	-1.995& 						&	-0.364 	&	-1.343&						&	-0.245 	&	6.658&\textsuperscript{***} 	&	1.216 \\
\bf Home Camera & \bf Weather station		&	\bf -1.354& 					& \bf -0.247 	&	\bf 1.468& 					& \bf 0.268 	&	\bf -1.313& 					& \bf -0.240 	&	\bf -0.606& 					& \bf -0.111 	&	\bf -0.262&  					& \bf -0.048 	& \bf 7.574&\textsuperscript{***} 	& \bf  1.383 \\
Home Camera & 	Alarm system 			&	-2.667&\textsuperscript{*}	&	-0.487 	&	5.588&\textsuperscript{***} 	&	1.020	&	-1.429& 						&	-0.261 	&	-2.779&\textsuperscript{**} 	&	-0.507 	&	-1.109& 						&	-0.202 	&	-1.481& 						&	-0.270 \\

    \cmidrule(r){1-2}
    \cmidrule(r){3-5}
    \cmidrule(r){6-8}
    \cmidrule(r){9-11}
    \cmidrule(r){12-14}
    \cmidrule(r){15-17}
    \cmidrule(){18-20}

\bf  Smoke detector & \bf  Thermostat 		&	\bf -0.308& 					& \bf -0.056 	& \bf 0.125& 					& \bf 0.023 	&	\bf -0.822& 					& \bf -0.150 	&	\bf 0.021& 					& \bf 0.004 	&	\bf -1.249& 					& \bf -0.228 	&	\bf 2.088&\textsuperscript{*} 	& \bf 0.381 \\ 
\bf  Smoke detector & \bf  Vacuum robot 	&	\bf 1.054& 					& \bf 0.193 	& \bf -2.009&					& \bf -0.367 &	\bf 0.436& 					& \bf 0.080 	&	\bf 1.160& 					& \bf 0.212 	&	\bf 0.787& 					& \bf 0.144 	&	\bf 3.288&\textsuperscript{**}	& \bf 0.600 \\
Smoke detector & Weather station 			&	2.234&\textsuperscript{*}		&	0.408 	&	-3.504&\textsuperscript{**} 	&	-0.640	&	1.797& 						&	0.328 	&	2.135&\textsuperscript{*}		&	0.390 	&	1.469& 						&	0.268 	&	5.129&\textsuperscript{***} 	&	0.936 \\
Smoke detector & Alarm system 			&	1.325& 						&	0.242 	&	0.000& 						&	0.000 	&	3.005&\textsuperscript{**} 	&	0.549 	&	0.923& 						&	0.169 	&	0.983 &						&	0.180 	&	-6.090&\textsuperscript{***} 	&	-1.112 \\

    \cmidrule(r){1-2}
    \cmidrule(r){3-5}
    \cmidrule(r){6-8}
    \cmidrule(r){9-11}
    \cmidrule(r){12-14}
    \cmidrule(r){15-17}
    \cmidrule(){18-20}

Thermostat & Vacuum robot 				&	1.498& 						&	0.274 	&	-2.067&\textsuperscript{*}	&	-0.377 	&	1.081& 						&	0.197	&	1.347& 						&	0.246 	&	2.062&\textsuperscript{*}		&	0.376 	&	2.418 &\textsuperscript{*}		&	0.441 \\
Thermostat & Weather station 				&	2.592&\textsuperscript{*}		&	0.473 	&	-4.089&\textsuperscript{***} 	&	-0.747 	&	2.445&\textsuperscript{*}		&	0.446 	&	2.466&\textsuperscript{*}		&	0.450 	&	2.838&\textsuperscript{**} 	&	0.518 	&	3.878&\textsuperscript{**} 	&	0.708 \\
Thermostat & Alarm system 				&	1.654& 						&	0.302 	&	-0.121& 						&	-0.022 	&	3.332&\textsuperscript{**} 	&	0.608 	&	1.000& 						&	0.182 	&	2.316&\textsuperscript{*}		&	0.423 	&	-6.361&\textsuperscript{***} 	&	-1.161 \\

    \cmidrule(r){1-2}
    \cmidrule(r){3-5}
    \cmidrule(r){6-8}
    \cmidrule(r){9-11}
    \cmidrule(r){12-14}
    \cmidrule(r){15-17}
    \cmidrule(){18-20}

Vacuum robot & Weather station 			&	1.020& 						&	0.186 	&	-1.948& 						&	-0.356	&	1.286& 						&	0.235 	&	1.025& 						&	0.187	&	0.712& 						&	0.130 	&	0.664& 						&	0.121 \\
Vacuum robot & Alarm system 				&	-0.038&						&	-0.007 	&	2.184&\textsuperscript{*}		&	0.399 	&	2.117&\textsuperscript{*}		&	0.387 	&	-0.539&						&	-0.098 	&	0.205& 						&	0.037 	&	-7.084&\textsuperscript{***} 	&	-1.293 \\    

    \cmidrule(r){1-2}
    \cmidrule(r){3-5}
    \cmidrule(r){6-8}
    \cmidrule(r){9-11}
    \cmidrule(r){12-14}
    \cmidrule(r){15-17}
    \cmidrule(){18-20}

Weather station & Alarm system 			&	-1.282& 						&	-0.234 	&	3.964&\textsuperscript{***} 	&	0.724 	& 	0.477 &						&	0.087 	&	-1.829& 						&	-0.334 	&	-0.851& 						&	-0.155	&	-8.227&\textsuperscript{***}	& 	-1.502 \\	

    \bottomrule
  \end{tabular}
     \begin{tablenotes}
      \footnotesize
      \item We are looking for pairs of product categories that are statistically significantly different in their security risk perception (criterion C1) and not different in other factors (criterion C2). Highlighted pairs of product categories fulfill these criteria. 
	\item Although we run multiple $t$-tests on the same data set for criterion C2, we did not perform a Bonferroni correction, because omitting the Bonferroni correction results in a conservative testing procedure for measuring the similarity of product categories.
	 %due to our interest in whether product categories are not different from each other. In fact, 
	  % Due to our interest in whether the product categories are not different from each other concerning the control factors (criterion C1), 
%(Because we wanted to find statistically significant differences between the product categories in regard of the perceived security-criticalness, we adjusted alphas for these tests.)
      \item Notation: Statistically significant with \textsuperscript{*}$p < 0.05$, \textsuperscript{**}$p < 0.01$, \textsuperscript{***}$p < 0.001$
  \end{tablenotes}
\end{table*}

\end{document}